\documentclass[a4paper,11pt]{article}
\usepackage{jcappub} 
\usepackage[T1]{fontenc} 
\usepackage{color}
\usepackage[update,prepend]{epstopdf}
\usepackage[toc,page]{appendix} 
\usepackage{pgffor} 
\usepackage{siunitx}
\usepackage{subfigure, wrapfig}

\def\nuebar{{\bar{\nu}_e}}

\def\LSD{liquid scintillator detectors}

\newcommand*\slfrac[2]{\left.#1\middle/#2\right.}

\title{\boldmath Prompt directional detection of galactic supernova by combining large liquid scintillator neutrino detectors}

\author[a]{V.\ Fischer,}
\author[a]{T.\ Chirac,}
\author[a,b,c]{T.\ Lasserre,}
\author[b]{C.\ Volpe,}
\author[a,b]{M.\ Cribier,}
\author[a]{M.\ Durero,}
\author[a]{J.\ Gaffiot,}
\author[a,b]{T.\ Houdy,}
\author[a]{A.\ Letourneau,}
\author[a]{G.\ Mention,}
\author[a]{M.\ Pequignot,}
\author[a]{V.\ Sibille,}
\author[a]{M.\ Vivier}

\affiliation[a]{Commissariat a l'\'energie atomique et aux \'energies alternatives, Centre de Saclay, IRFU, 91191 Gif-sur-Yvette, France}
\affiliation[b]{Astroparticules et Cosmologie APC, 10 rue Alice Domon et L\'eonie Duquet, 75205 Paris cedex 13, France}
\affiliation[c]{Institute for Advanced Study, Technische Universit\"{a}t M\"{u}nchen, Lichtenbergstrasse 2 a, D-85748 Garching, Germany.}

\emailAdd{vincent.fischer@cea.fr}
\emailAdd{tchirac@gmail.fr}
\emailAdd{thierry.lasserre@cea.fr}
\emailAdd{volpe@apc.univ-paris7.fr}
\emailAdd{michel.cribier@cea.fr}
\emailAdd{mathieu.durero@cea.fr}
\emailAdd{jonathan.gaffiot@cea.fr}
\emailAdd{thibaut.houdy@cea.fr}
\emailAdd{alain.letourneau@cea.fr}
\emailAdd{guillaume.mention@cea.fr}
\emailAdd{maxime.pequignot@cea.fr}
\emailAdd{valerian.sibille@cea.fr}
\emailAdd{matthieu.vivier@cea.fr}

\abstract{
Core-collapse supernovae produce an intense burst of electron antineutrinos in the few-tens-of-MeV range. Several Large Liquid Scintillator-based Detectors (LLSD) are currently operated worldwide, being very effective for low energy antineutrino detection through the Inverse Beta Decay (IBD) process. In this article, we develop a procedure for the prompt extraction of the supernova location by revisiting the details of IBD kinematics over the broad energy range of supernova neutrinos. Combining all current scintillator-based detector, we show that one can locate a canonical supernova at 10~kpc with an accuracy of~45~degrees (68\% C.L.). After the addition of the next generation of scintillator-based detectors, the accuracy could reach~12~degrees (68\% C.L.), therefore reaching the performances of the large water \v{C}erenkov neutrino detectors. We also discuss a possible improvement of the SuperNova Early Warning System (SNEWS) inter-experiment network with the implementation of a directionality information in each experiment. Finally, we discuss the possibility to constrain the neutrino energy spectrum as well as the mass of the newly born neutron star with the LLSD data. 
}

\begin{document}
\maketitle

\section{Introduction}

Neutrinos\footnote{For simplicity reasons and unless stated otherwise, the term neutrino will be used for both electron neutrinos and electron antineutrinos in the following.} have already been detected from the type~II supernova SN1987A that appeared in the Large Magellanic Cloud (LMC) in 1987. At that time three detectors, Kamiokande~II~\citep{Hirata:1987hu}, IMB~\citep{Bionta:1987qt} and Baksan~\citep{Alekseev:1988gp} detected twenty-nine neutrino events, hence opening the field of extrasolar neutrino astronomy (see Ref.~\citep{Vissani:2014doa} for a recent review). 
	 	 
Since they leave the dense medium of the core-collapse supernova at the early stages of the process, neutrinos precede the optical photons by several hours~\citep{Longo:1987ub}. An early pointing could allow the astronomical community to quickly locate the supernova and consequently study the explosion early phases with the world's telescope armada. 
	 	 
The estimated core-collapse supernova rate in our Galaxy is a few per century. The possible progenitor could be located anywhere, nevertheless the supernova probability distribution computed from the red giant stars distribution is centered around 11.9~kpc with a RMS dispersion of 6~kpc~\citep{Mirizzi:2006xx}. 
	 	 
Neutrino scattering on electrons in a very large water \v{C}erenkov detector, such as Super-Kamiokande~\citep{Fukuda:2002uc}, provides the best way to locate a supernova through neutrinos, with an accuracy of about 8 degrees at 10~kpc (95\% C.L.)~\citep{Scholberg:2011kt}. However, the pointing accuracy is degraded by the IBD reactions, with a neutrino scattering signal over IBD noise ratio of about 1/30. On the other hand, the large cross section of the IBD reaction compared to neutrino scattering allows the detection of supernova $\nuebar$ in smaller detectors.
	 	 
In this work, we revisit the early pointing of a core-collapse supernova using the IBD reaction in LLSD. Although it does not provide an event by event pointing ability, the high cross section of this reaction leads to a high number of interactions, almost free of backgrounds. The statistical analysis of the angular distributions of the reaction products (positron and neutron) can be correlated to the neutrino direction and its associated uncertainty, enabling an early pointing of the supernova.

While a single experiment can independently provide a most probable supernova location, we show that the prompt combination of data from all large liquid scintillator detectors could reduce the field of view for further optical-based astronomical observations. We conclude by discussing other astrophysical information that can be retrieved such as the mass of the newly born neutron star and the supernova distance.

\section{Core collapse supernova neutrino signal}
\label{sec:SN_neutrino_signature}

In this section we outline the physics of core collapse supernovae and the associated neutrino signals. 
The present description is by no means complete and further details can be found in Ref.~\citep{Janka:2012wk} for instance.

Core-collapse supernovae are massive stars that undergo a gravitational collapse at the end of their life, emitting 99$\%$ of the gravitational energy through neutrinos and antineutrinos of all flavors. A supernova neutrino luminosity curve closely follows the different phases of the explosion, from the tens-of-milliseconds burst of $\nu_e$ (due to electron capture on nuclei), to the hundreds of milliseconds accretion phase and the ten-second-long cooling phase of the left proto-neutron star. During accretion and cooling, neutrinos and antineutrinos of all flavors are emitted with $\langle E_{\nu_e} \rangle < \langle E_{\bar{\nu}_e} \rangle < \langle E_{\nu_x} \rangle $ with $\nu_x = \nu_{\mu}, \nu_{\tau},
\bar{\nu}_{\mu}, \bar{\nu}_{\tau}$. This hierarchy of average energies reflects neutrino interactions with matter at tree level, and the fact that the star is neutron-rich. 
At the neutrinosphere boundary, where neutrinos start free-streaming, the neutrino fluxes are well described by a Fermi-Dirac or a power-law distribution~\cite{Keil:2002in}. 
Outside the neutrinosphere, the Mikheev-Smirnov-Wolfenstein (MSW) effect~\citep{Wolfenstein:1977ue,Mikheev:1986gs} is well established. Additional flavor conversion phenomena occur due to neutrino self-interaction as well as the presence of shock waves and turbulence.
While many features concerning the mechanisms and the occurrence of such flavor conversion phenomena have been unraveled, important open questions remain~\cite{Volpe:2013kxa}. 
Note also that the recently discovered lepton-number emission self-sustained asymmetry (LESA) is expected to produce a dipole pattern during the accretion phase~\cite{Tamborra:2014aua}. In fact, as we will show, the shape of the neutrino spectra has a negligible impact on the directional reconstruction of a core-collapse supernova. 

\begin{figure}[h!]
\begin{center}
\includegraphics[width = 0.55\textwidth]{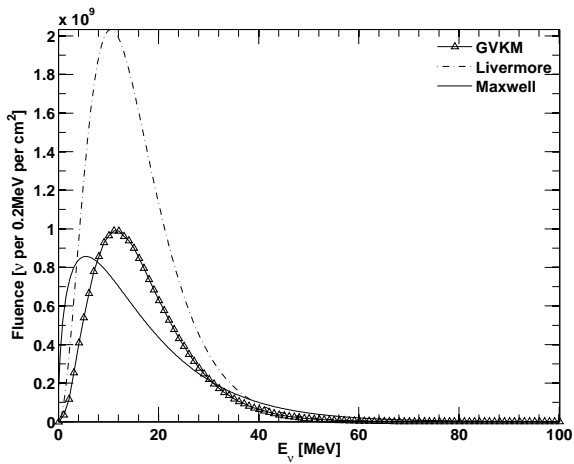}
\caption{Expected core-collapse supernova $\nuebar$ emission spectra (integrated over 10~s) based on the Maxwell, Livermore, and GKVM fluxes at 10~kpc. The three spectra are normalized to the same energy released by the supernova, $3 \times 10^{52}$~ergs ($2 \times 10^{58}$~MeV), equally distributed among each neutrino and antineutrino species. The integrated fluences of the three distributions are respectively $9.7 \times 10^{10}$~$\nuebar$.cm$^{-2}$, $1.9 \times 10^{11}$~$\nuebar$.cm$^{-2}$ and $9.7 \times 10^{10}$~$\nuebar$.cm$^{-2}$. }
\label{fig:SNflux}
\end{center}
\end{figure}

In what follows we focus on the $\nuebar$ emission, relevant for IBD detection. We are considering three examples of neutrino flux prediction: a maxwellian-like distribution, the Livermore flux~\citep{Totani:1997j}, and the GVKM (Gava-Kneller-Volpe-McLaughlin) flux~\citep{Gava:2009pj}. Figure~\ref{fig:SNflux} displays the three expected energy spectra. We will show that the shape of these distributions only has a negligible effect on our following results.\\

The maxwellian distribution is a common thermal distribution used to simply describe the energy spectrum of the neutrinos immediately after the burst. The Livermore distribution, similar to a Fermi-Dirac or a power-law distribution, describes the $\nuebar$ energy spectrum after exiting the supernova neutrinosphere, without conversion or oscillation effects.
The GVKM distribution provides a modeling of flavor conversion due to neutrino self-interactions and shockwave effects that affect the fluence and the shape of the $\nuebar$ spectrum. Details of these effects can be found in~\citep{Gava:2009pj}. It reduces the number of expected electronic antineutrino events by $\sim$50\% with respect to the Livermore flux. It is an improved calculation of the expected energy spectrum and we therefore consider it as our reference flux. 


\section{Neutrino detection in large liquid scintillator detectors}
\label{sec:LLSD_detection}

\subsection{Neutrino detection through IBD}
\label{sec:SN_detection_IBD}
Core-collapse supernova neutrinos can interact through charged-current reactions on nuclear targets. Here we focus on the IBD in which only $\nuebar$ participate, since it leads to the highest number of events in liquid scintillator detectors. 
Electronic antineutrinos are detected via IBD on free protons: $\nuebar + p \to e^+ + n$ for $E_{\nuebar} > E_{\rm thr} \simeq 1.806$~MeV in the laboratory frame where the free proton is at rest. 

\begin{figure}[h!]
\begin{center}
\includegraphics[width = 0.45\textwidth]{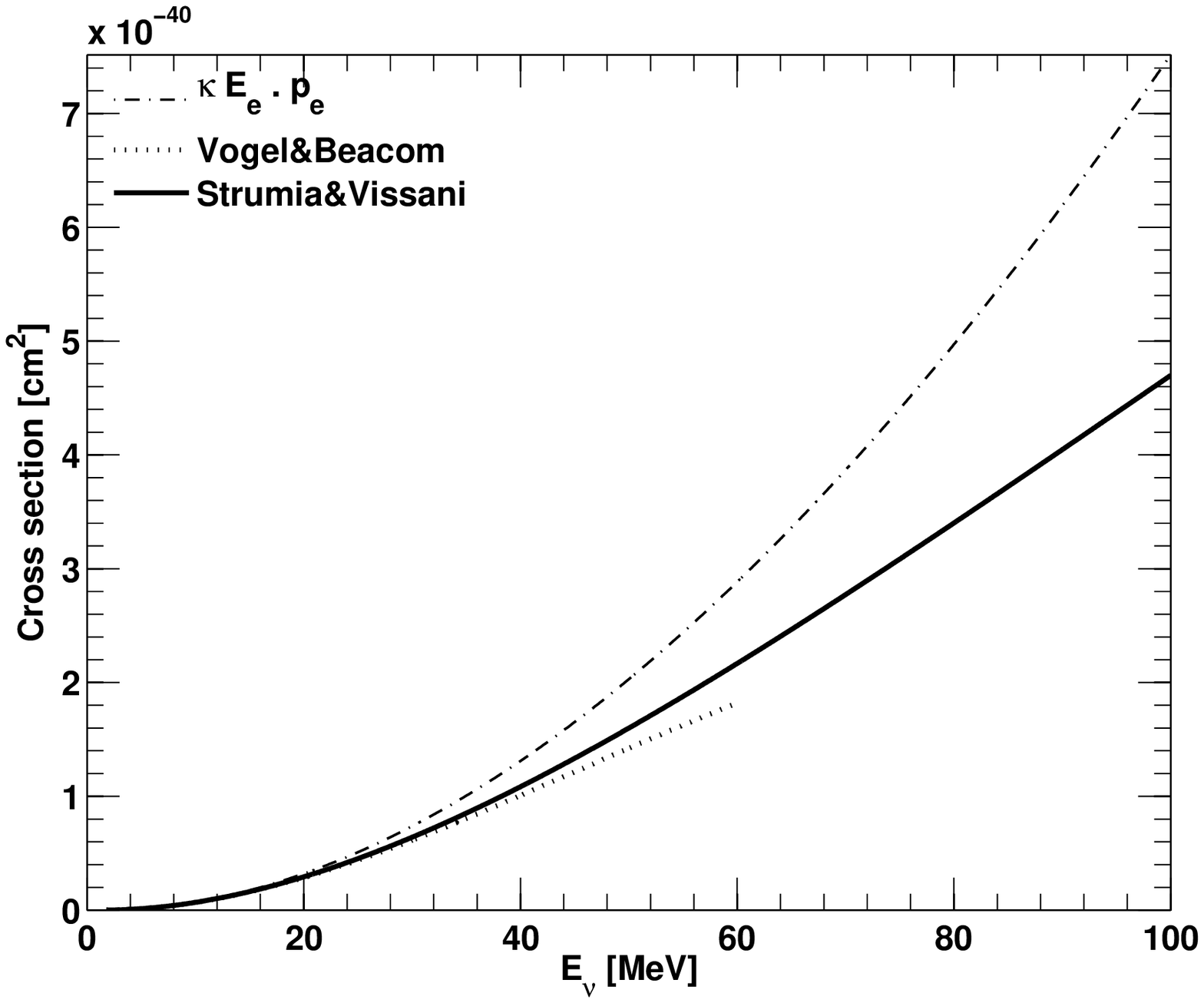}
\includegraphics[width = 0.51\textwidth]{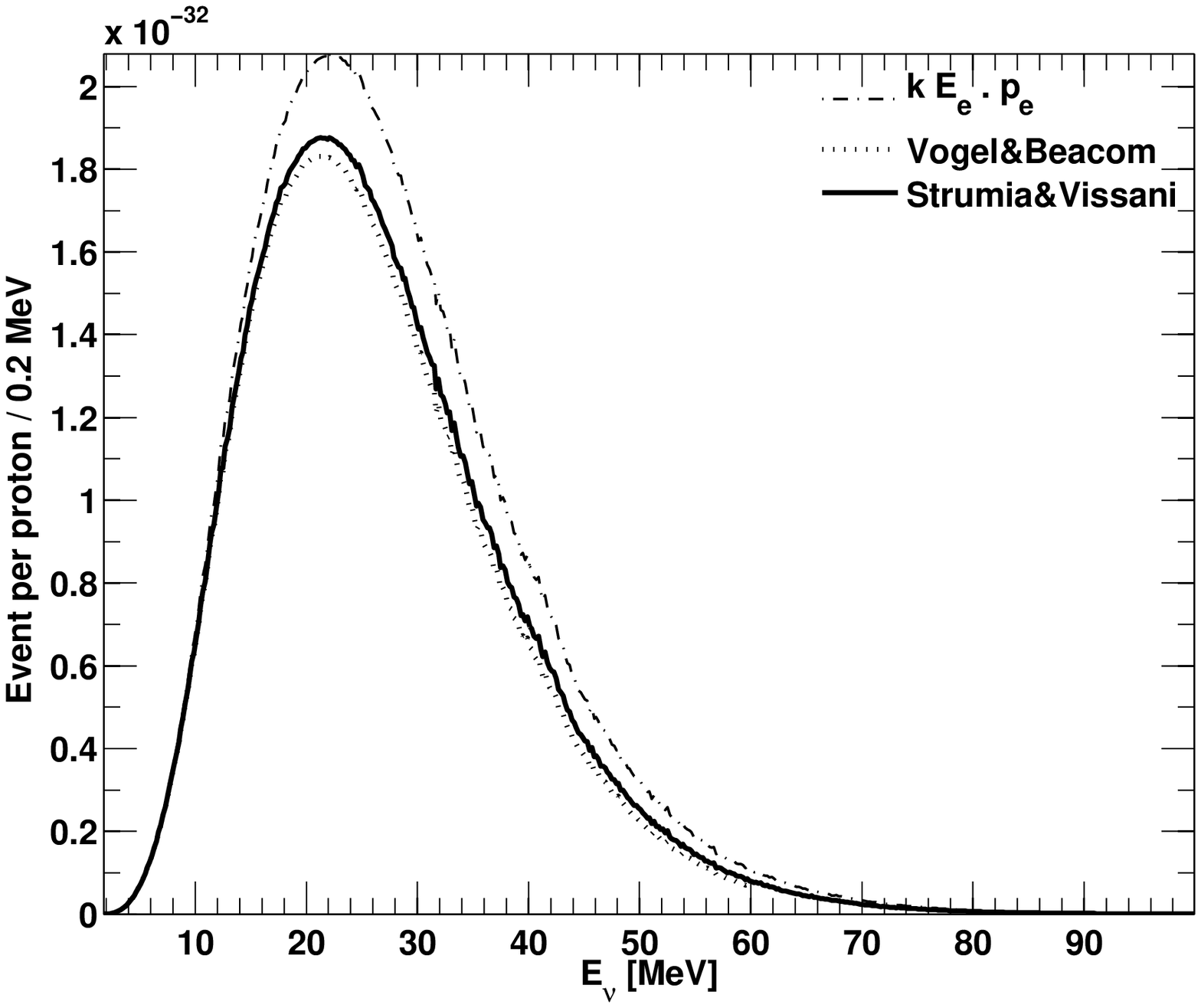}
\caption{Left: IBD cross section based on three different approximations that differ significantly for energies greater than 30 MeV. Right: Expected neutrino spectrum for a supernova located at 10~kpc, normalized to the interaction on one free proton.}
\label{fig:ibdrate}
\end{center}
\end{figure}

The IBD reaction cross section can be precisely computed from the \mbox{V-A}~theory of weak interaction. However, depending on the considered energy scale, some approximations are often performed to simplify the mathematical expressions. For core-collapse supernova $\nuebar$ energy range, up to $E_\nuebar$=100~MeV, we consider the cross section computed by A.~Strumia and F.~Vissani in Ref.~\citep{Strumia:2003zx} and displayed in Figure~\ref{fig:ibdrate}: 

\def\dxscndt{ \frac{G_F^2 \cos^2 \theta_C}{2\pi(s-m_p^2)^2}\vert{\cal{M}}\vert ^2}
\begin{equation}
\frac{\mathrm{d}\sigma}{\mathrm{d}t} =\dxscndt , 
\label{eqn:dsigmadt}
\end{equation}
where $G_F$ is the Fermi constant, $\cos\theta_C$ is the Cabbibo angle, $m_{p}$ is the proton mass, $\cal{M}$ is the invariant amplitude for the IBD reaction explicitly given in Ref.~\citep{Strumia:2003zx}, $s=(p_\nu+p_p)^2$ and $t=(p_\nu-p_e)^2$ are expressed as functions of the neutrino, proton and positron 4-momenta $p_\nu$, $p_p$ and $p_e$. 
Other cross section expressions such as a naive ``p$_{\text{e}}$E$_{\text{e}}$'' relation (see Ref.~\citep{Strumia:2003zx}) and computations carried out by P.~Vogel and J.~Beacom~\citep{Vogel:1999zy} are displayed in Figure~\ref{fig:ibdrate} as well. As stated in Ref.~\citep{Vogel:1999zy}, the total cross section computed by Vogel-Beacom is only valid up to 60~MeV.

Following the computation of Ref.~\citep{Strumia:2003zx} the outgoing positron and incoming $\nuebar$ energies are related by:
\begin{equation}
E_e = \frac{ \left( E_{\nu} - \delta \right) \left( 1 + \epsilon \right) + \epsilon \cos{\theta_{e}} \sqrt{\left( E_{\nu} - \delta \right)^{2} - m_{e}^{2} \kappa} }{ \kappa },
\label{eqn:Ee2Enu}
\end{equation}
with
\begin{eqnarray}
\delta =& &\left( m_{n}^{2} - m_{p}^{2} - m_{e}^{2} \right) / 2 m_{p} \\
\epsilon =& &E_{\nu} / m_{p} \\
\kappa =& &\left( 1 + \epsilon \right)^{2} - \left( \epsilon \cos{\theta_{e}} \right)^{2},
\end{eqnarray}
where $E_\nu$ is the neutrino energy, $E_e$ is the outcoming positron total energy, $\theta_{e}$ is the angle between the positron and the incoming antineutrino directions in the laboratory frame and $m_{n}$, $m_{p}$ and $m_{e}$ are respectively the neutron, proton and electron masses. 
We call \textit{visible energy} the positron energy deposited in the detector by ionization and annihilation, corresponding to $E_{\rm vis} = E_{e} + m_{e}$. 

After its emission, the neutron undergoes several elastic scatterings in the hydrogen-rich medium of the liquid scintillator. This process called \textit{moderation} decreases its kinetic energy until it reaches typical thermal equilibrium at $\sim$0.025~eV. After thermalization, the neutron diffuses before its capture on a nucleus. The average duration of this diffusion process depends on the liquid composition. It varies from tens of microseconds for a Gadolinium(Gd)-doped liquid to hundreds of microseconds for a non-doped liquid. The use of Gd doping increases the detection efficiency while reducing the impact of accidental background. However, it is worth noting that despite the Gd-doping, a fraction of the neutrons will be captured on hydrogen (H), depending on the liquid composition. The current technologies developed for the recent reactor neutrino experiments~\citep{Abe:2012tg,An:2013uza,Ahn:2010vy} guarantee the possibility of doping the next generation of large detectors with Gd at the multikiloton-scale. 

An IBD event in a LLSD is thus characterized by a prompt positron event which deposits a visible energy between 1 and up to 100~MeV, followed by a delayed burst of $\gamma$-rays arising from neutron capture on a target nucleus within $\tau \sim 10-1000\;\mu$s, depending of the target liquid composition. These prompt and delayed events are spatially correlated, within a few cubic meters.

\begin{figure}[h!]
\begin{center}
\includegraphics[width = 0.56\textwidth]{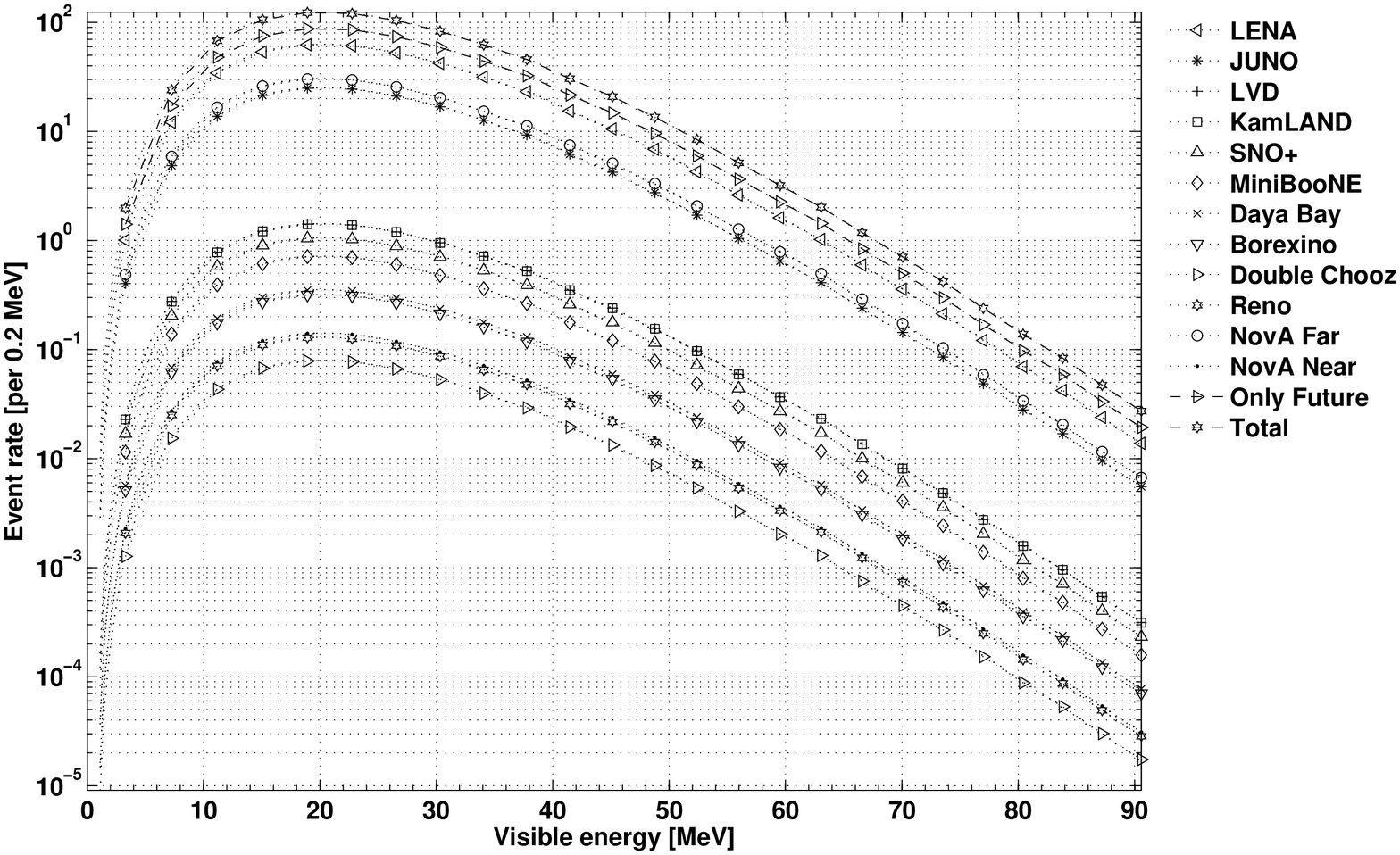}
\includegraphics[width = 0.43\textwidth]{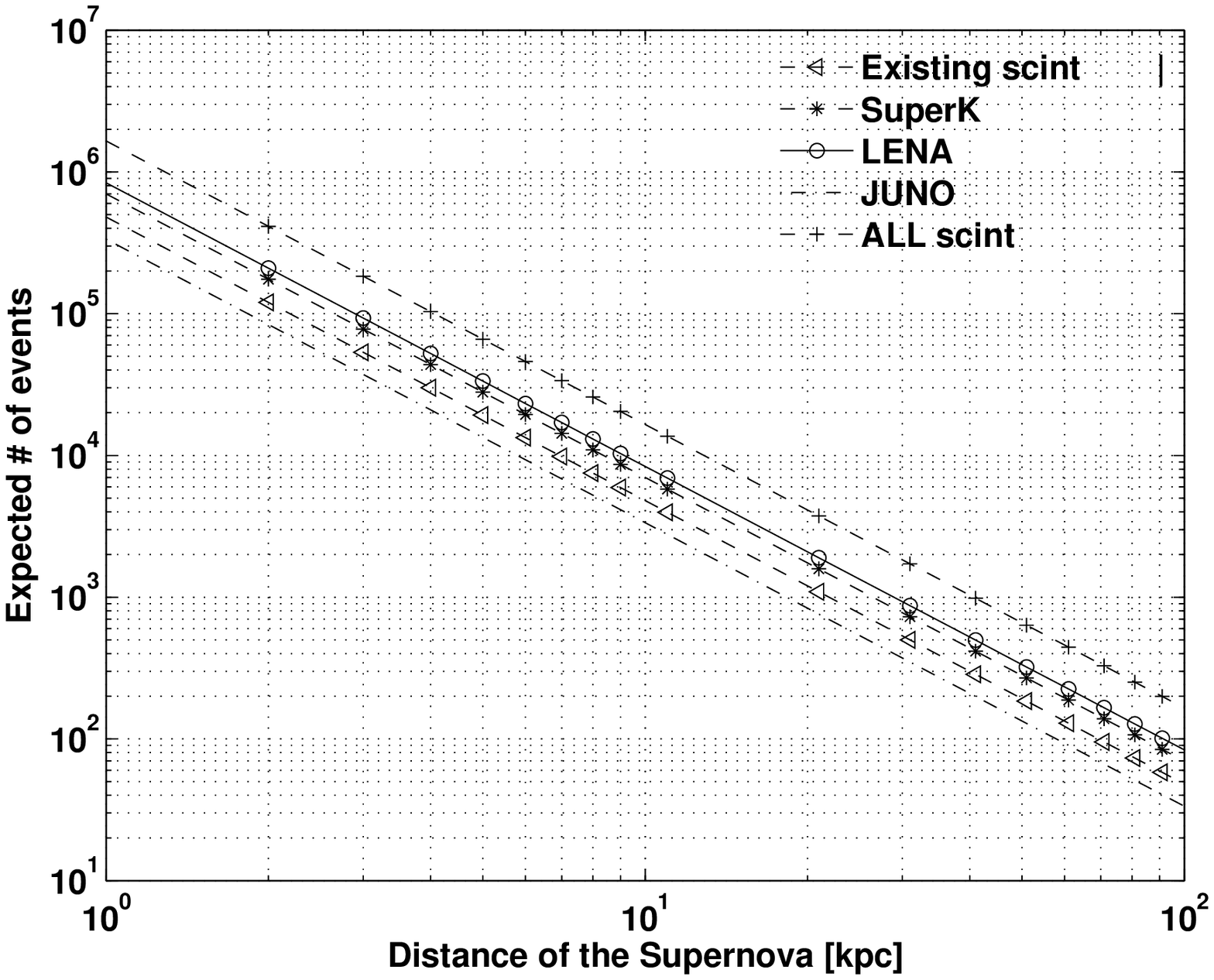}
\caption{Left: IBD interaction rate in various LLSD as a function of the visible energy, for a core collapse supernova located at 10~kpc. ``Future only'' and ``All'' are respectively the combination of LENA and JUNO and the combination of all LLSD, current and future. Right: IBD interaction rate for LLSD as a function of the supernova distance. The number of expected IBD events in Super-Kamiokande is shown for comparison.}
\label{fig:SNrate}
\end{center}
\end{figure}

\subsection{Running detectors}
\label{subsec:running_detectors}

The detection of supernova neutrinos in liquid scintillators relies on the existence of numerous large scale detectors with a mass bigger than several tens of tons. These volumes are large enough to detect a significant number of events and tag a supernova in our Galaxy or its closest satellites, such as the small and large Magellanic Clouds, with a negligible fake detection rate, thanks to the coincidence between detectors.
The non-exhaustive list of detectors considered in this work and their relevant characteristics are collected in Table~\ref{table:Detectors}. Their respective expected event rates are displayed on Figure~\ref{fig:SNrate} and Table~\ref{SuperNustradamus}.

\begin{table}[h!]
\caption{Main detector features used in our simulation.}
 \centering 
{\footnotesize
\begin{tabular*}{\textwidth}{@{\extracolsep{\fill}}c c c c c c c c c}
 \hline
Detector	&	Dimensions (m)				&Shape 		& V (m$^3$) 	&	free H/m$^3$ 	& $\#$ free H\\
\hline
LVD		&	a=10, b=13.2, c=22.7 			&Box 		&1020		&$7.46\ .10^{28}$	&$7.6\ .10^{31}$\\
KamLAND		&	r=6.50					&Sphere		&1150 		&$6.60\ .10^{28}$	&$7.6\ .10^{31}$\\
SNO+		&	r=6.00					&Sphere		&904		&$6.24\ .10^{28}$	&$5.6\ .10^{31}$\\
Borexino	&	r=4.25					&Sphere		&321		&$5.30\ .10^{28}$	&$1.7\ .10^{31}$\\
Daya Bay	&	r=2.00, h=4.00				&Cylinder	&50($\times$6)		&$6.24\ .10^{28}$	&$1.8\ .10^{31}$\\
Double Chooz	&	r=1.70, h=3.55				&Cylinder	&32($\times$2)		&$6.55\ .10^{28}$	&$0.4\ .10^{31}$\\
Reno		&	r=2.00, h=4.40				&Cylinder	&55($\times$2)		&$6.24\ .10^{28}$	&$0.7\ .10^{31}$\\
NOvA (far)	&	a=15.7, b=15.7,c=132			&Box		&32500		&$6.24\ .10^{28}$	&$2.0\ .10^{33}$\\
NOvA (near)	&	a=3.5, b=4.8, c=9.58			&Box		&161		&$6.24\ .10^{28}$	&$1.0\ .10^{31}$\\
MiniBooNE	&	r=5					&Sphere		&523		&$7.31\ .10^{28}$	&$3.8\ .10^{31}$\\
\hline
\label{table:Detectors}
\end{tabular*} 
}
\end{table}

NO$\nu$A (NuMI Off-Axis $\nuebar$ Appearance) is an experiment dedicated to observe $\nu_{e}$ appearance using Fermilab's $\nu_{\mu}$ beam NuMI~\citep{Ayres:2004js}. It operates above ground with two detectors, each made of individual cells of liquid scintillator of 3.87~cm $\times$ 6.00~cm $\times$ 15.7~m. With 12,480~cells, NO$\nu$A's near detector consists of 145~tons of active material while the far detector and its 761'856~cells encloses 24~ktons of liquid scintillator. Its important background rate at sub-GeV energies and its energy threshold, above the neutron capture on H at 2.2~MeV, does not allow an efficient neutrino burst detection nor a neutron vertex reconstruction yet. NO$\nu$A will thus not be taken into account in the following direction reconstruction analyses.

The Large Volume Detector (LVD)\citep{Alberini:1986ew} is located at the Laboratori Nazionali del Gran Sasso (42$^{\circ}$ 27' 10" N, 13$^{\circ}$ 34' 30" E), at the depth of 3500~m.w.e.. It is a 1~kt liquid scintillator detector in the form of an array of 840~scintillator counters, 1.5~m$^{3}$ each. While its active mass allows the detection of an important number of neutrinos, the size of LVD's individual scintillating counters prevents any precise position reconstruction. Hence, like NO$\nu$A, it will not be taken into account in the following directionality analyses.

From now on, the following detectors will be considered in our directionality analyses.\\
The KamLAND experiment \citep{Eguchi:2002dm} is a 1~kton liquid scintillator (80\% dodecane and 20\% pseudocumene) balloon located within the Kamioka mine in Japan (36$^{\circ}$ 25' N, 137$^{\circ}$ 18' E), at 2700~mwe under the Ikenoyama mountain. 

The SNO+ experiment \citep{Chen:2008un} is currently in its final construction phase at the SNOLAB laboratory (46$^{\circ}$ 28' 30" N, 81$^{\circ}$ 12' 04"), shielded with an overburden of 6000~m.w.e.. SNO+ will be filled with 780~tons of Linear Alkyl Benzene (LAB).

The Borexino experiment \citep{Alimonti:2008gc} is located at the Laboratori Nazionali del Gran Sasso in Italy (42$^{\circ}$ 27' 10" N, 13$^{\circ}$ 34' 30" E) and shielded with an overburden of 3500~m.w.e.. It consists of 278~tons of pseudocumene-based liquid scintillator.

Three reactor neutrino oscillation experiments, Daya Bay~\citep{An:2013uza}, Double Chooz~\citep{Abe:2012tg}, and RENO~\citep{Ahn:2010vy} are currently taking data to measure the $\theta_{13}$ mixing angle. They all consists of several identical detectors located within two kilometers from several powerful nuclear power plant cores used as intense electron antineutrino sources. 
The Daya~Bay experiment is located in the Guang-Dong Province, on the site of the Daya~Bay nuclear power station (22$^{\circ}$ 36' 59" N, 114$^{\circ}$ 32' 28.1" E). Each of the six cylindrical identical detector modules contains an effective volume of 20~tons of 1~g/l Gd-loaded liquid scintillator, and 22 tons of non-doped scintillator.
The Double~Chooz experiment is located close to the twin reactor cores of the Chooz nuclear power station located in the French Ardennes (50$^{\circ}$ 08' 43" N, 4$^{\circ}$ 80' 165" E). The two cylindrical identical detectors contain a 8-ton fiducial volume of liquid scintillator doped with 1~g/l of Gd and 18 tons of non-doped scintillator.
The RENO experiment is located on the site of the Yonggwang nuclear power plant in Korea, about 400~km south of Seoul (35$^{\circ}$ 24' 04" N, 126$^{\circ}$ 25' 31.5" E). The two cylindrical identical detectors contain a 16-ton fiducial volume of liquid scintillator doped with 1~g/l of Gd and 39~tons of non-doped scintillator. Given the large volume fraction of undoped scintillator liquid in those three detectors, in the following we only consider the events expected from neutron capture on H. Considering the energy and vertex reconstruction enhancement brought by neutron capture on Gd, this is a conservative approach.

The MiniBooNE detector \citep{Bazarko:1999hq}, located at Fermilab (41$^{\circ}$ 83' N, -88$^{\circ}$ 26' E), is a spherical volume containing 680 tons of mineral oil built at ground level. 
The project of adding PPO in the mineral oil proposed in Ref.~\citep{Dharmapalan:2013zcy} indicates the detector might still be running for a few years. The addition of PPO would enhance scintillation and set MiniBooNE in our LLSD list.

\begin{table}[h!]
 \caption{Expected number of interactions taking into account the exact liquid scintillator chemical compositions and densities, for a 10~kpc core-collapse supernova.}
 \centering 
{\footnotesize
\begin{tabular*}{0.5\textwidth}{@{\extracolsep{\fill}}ccccccc}
 \hline
 &  Livermore & GVKM\\
\hline
LVD & 319 &		190\\
KamLAND &318&    189\\
SNO+ &237 &141\\
Borexino & 71 &43\\
Daya Bay &79&	47\\
Double Chooz &18& 11\\
RENO &29& 17\\
NO$\nu$A far & 6811 & 4057 \\
NO$\nu$A near & 32 & 19 \\
MiniBooNE &161&		96\\
\hline
\end{tabular*}
\label{SuperNustradamus}
}
\end{table}

\subsection{Forthcoming detectors}
\label{subsec:future_detectors}

Beyond the currently running detectors two very large \LSD\ projects are being considered: JUNO and LENA. These detectors, besides providing an answer to some of the major open questions remaining in the neutrino field, are perfect apparatus for studying core-collapse supernovae. 

JUNO (Jiangmen Underground Neutrino Observatory) is under construction and will consist of a 20~kt LAB-based liquid scintillator detector located at Jiangmen (22$^{\circ}$37' N, 112$^{\circ}$70' E) in South China \citep{Zhan:2013cxa}. Being 20 times as massive as SNO+, the same factor on the expected event number is expected.

LENA (Low Energy Neutrino Astronomy) is a project designed to be a next-generation liquid scintillator detector on the scale of 50 kt~\citep{Wurm:2011zn}. Its location is still under investigation but the most probable is the Pyh\"asalmi mine (63$^{\circ}$39' N, 26$^{\circ}$2' E, 4000 m.w.e.), located in the Pyh\"aj\"arvi province in Finland.

\section{Supernova early pointing in LLSD}
\label{sec:SN_pointing}

Our goal is to determine the supernova direction in the sky from a statistical analysis of all interactions associated to the electron antineutrino burst, taking advantage of the IBD kinematic features keeping some memory of the incoming neutrino direction. For $\nuebar$ IBD interactions in LLSDs, prompt and delayed event features, such as energy and interaction vertices, are determined on an event-by-event basis in each detector. For each event, one can then define a vector starting at the prompt vertex and ending at the delayed vertex. This information is complemented by the prompt visible energy. 

\subsection{Relevant IBD kinematics features}
\label{subsec:IBDkinematics}
In this section we review the IBD relevant kinematic features used to constraint the $\nuebar$ incoming direction. We compare analytical calculations of Ref.~\citep{Vogel:1999zy} with our toy Monte Carlo simulation described in Appendix~\ref{section:toyMC}.

\subsubsection*{Visible energy}
\begin{figure}[h!]
\begin{center}
\includegraphics[width = 0.6\textwidth]{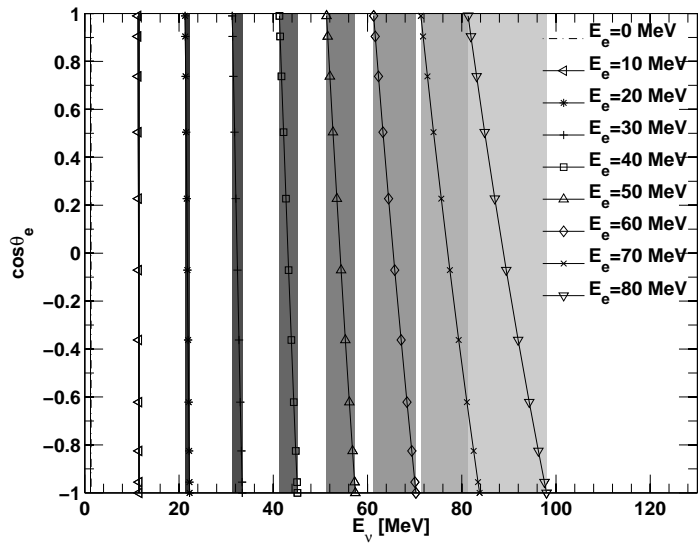}
\caption{Relationship between $E_\nu$ and $E_e$ as a function of the positron emission angle, $\cos{\theta_{e}}$. The grey bands correspond to the range of possible reconstructed neutrino energies for a given positron energy.}
\label{fig:compareEeEnu}
\end{center}
\end{figure}

The relation between the positron energy and the neutrino energy is straightforward at reactor neutrino energies, $E_\nu<$10~MeV, but it gets ambiguous at higher energies. Eq.~\ref{eqn:Ee2Enu} shows that $E_{\nu}$ both depends on $E_e$ and $\cos{\theta_{e}}$. Since only $E_e$ can be measured, a degeneracy appears in the $\nuebar$ energy estimate. For reactor neutrinos ($E_\nu<$10~MeV), this degeneracy is weak, as shown by Figure~\ref{fig:compareEeEnu}, while it strengthens at higher energies. Since $\cos{\theta_{e}}$ cannot be measured, this induces an error of up to 10\% in the $\nuebar$ energy reconstruction.
The positron takes the maximum of the energy when it is emitted along the neutrino direction. 
On the other hand, it goes preferably slightly backwards when taking a relatively small energy, the rest being available to the neutron that can therefore scatter further forwards with larger emission angles. 

\subsubsection*{Positron and neutron directions}
The outgoing positron angular distribution is obtained by multiplying Eq.~\ref{eqn:dsigmadt} with the following Jacobian:
\def\dtdcos{\frac{2 p_e E_\nu }{m_p+E_\nu \left(1-\slfrac{E_e}{p_e}\cos{\theta_{e}} \right)}}
\begin{equation}
\frac{\mathrm{d}t}{\mathrm{d}\cos{\theta_{e}}} = \dtdcos  .
\label{eqn:dtdcostheta}
\end{equation}

\begin{figure}[h!]
\begin{center}
\includegraphics[width = 0.49\textwidth]{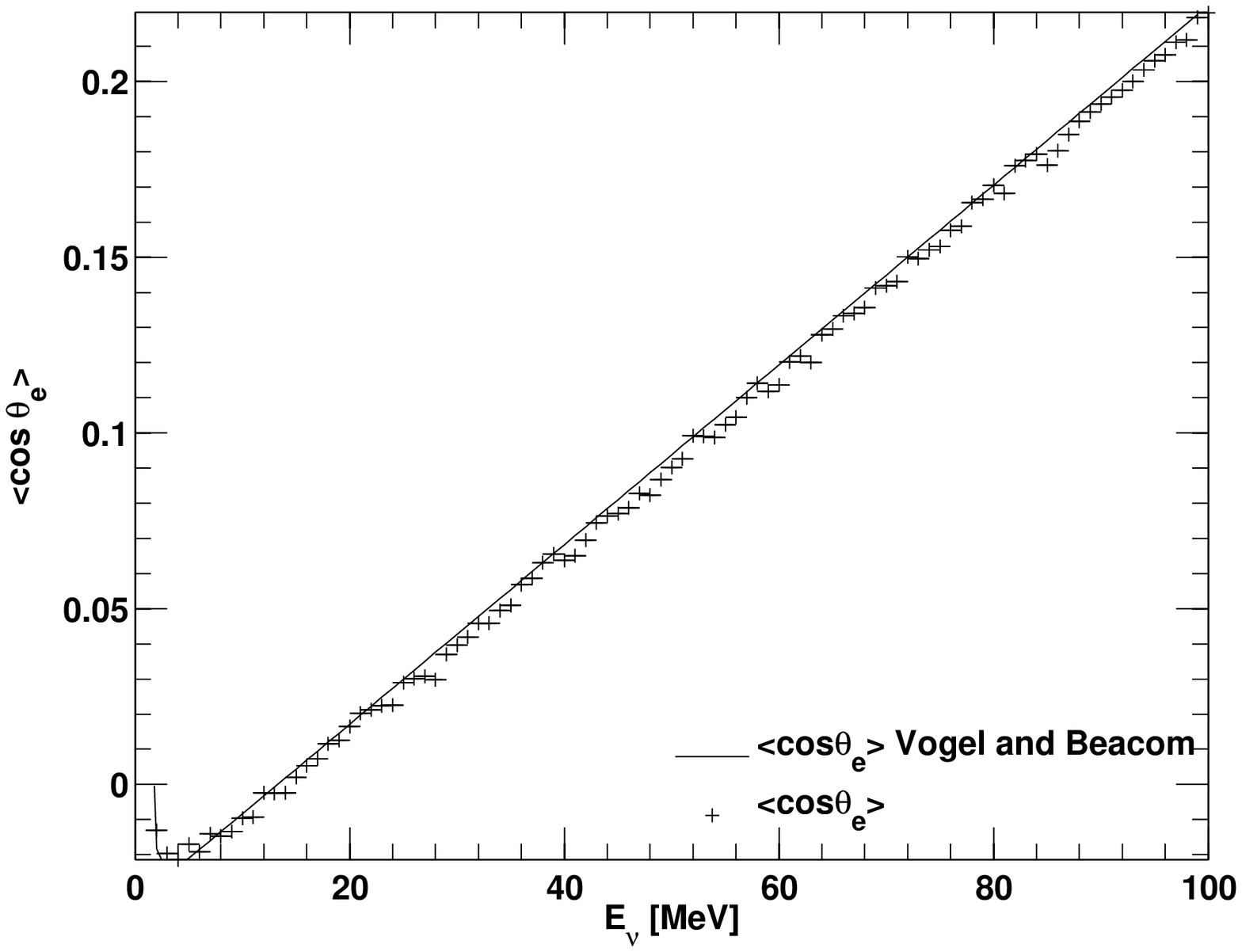}
\includegraphics[width = 0.49\textwidth]{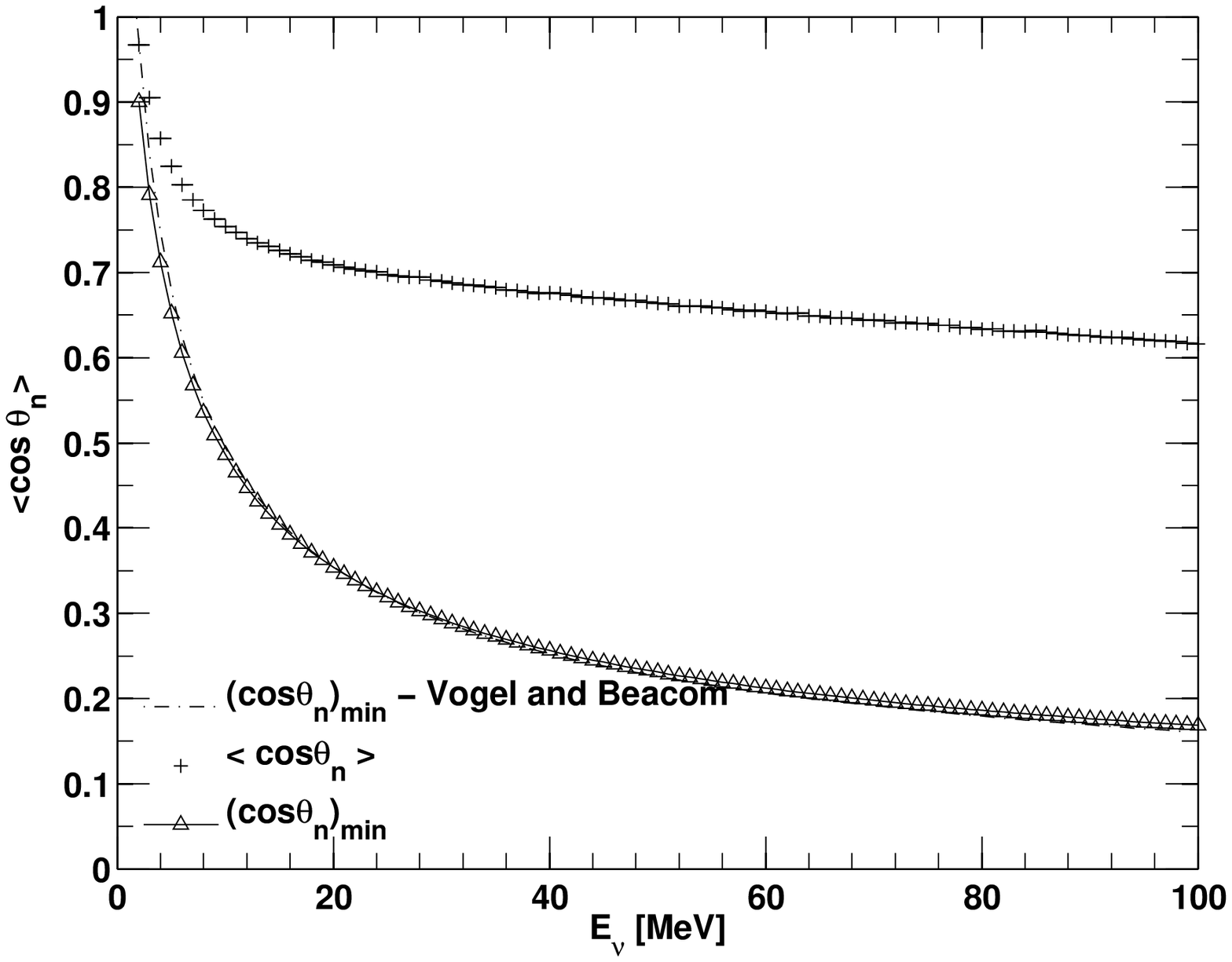}
\caption{Left: Average cosine of the outgoing positron angle, $\cos{\theta_{e}}$, as a function of neutrino energies between $E_{th}$ and 100~MeV. Right: Average cosine of the outgoing neutron angle, $\cos{\theta_{n}}$. 
Both panels show results of our toy Monte Carlo based on the Strumia-Vissani cross section compared to the analytical evaluations of Vogel-Beacom.}
\label{fig:cosTheta}
\end{center}
\end{figure}

For each pair ($E_\nu$,$\cos{\theta_{e}}$), $E_e$ is obtained from Eq.~\ref{eqn:Ee2Enu}. 
For $E_\nu < 15$~MeV, the positron is emitted on average in a slightly backward direction, whereas it moves preferably forward as the incident neutrino energy is higher, approximately following the form $<\cos{\theta_{e}}> \simeq -0.034 \slfrac{\sqrt{E_e^2-m_e^2}}{E_e}+2.4 \slfrac{E_\nu}{m_p}$ computed in Ref.~\citep{Vogel:1999zy}.
The neutron is always emitted in the forward hemisphere, with a maximum allowed angle increasing with the neutrino energy. Figure~\ref{fig:cosTheta} shows the mean values of $\cos{\theta_{e}}$ and $\cos{\theta_{n}}$ obtained from our toy Monte Carlo simulation compared with the Vogel-Beacom analytic computation of the differential cross section, valid up to $\sim$~150~MeV. The positron-neutron vector, our directionality observable, is thus energy dependent.

\subsubsection*{Positron and neutron mean paths}

After their creation both particles move into the detector with respect to the interaction vertex before being annihilated or captured.\\

As it crosses liquid scintillator, the positron loses its energy by ionization until it annihilates with an electron thus creating a pair of gamma rays. The mean path length between the positron creation at the interaction point and its annihilation obtained from our simulation is displayed in Figure~\ref{fig:ranges}, for different positron energies. As explained in Appendix~\ref{section:toyMC}, the path length must be corrected to account for the detector reconstruction since the track of deposited energy is reconstructed as a point-like vertex.

\begin{figure}[h!]
\begin{center}
\includegraphics[width = 0.7\textwidth]{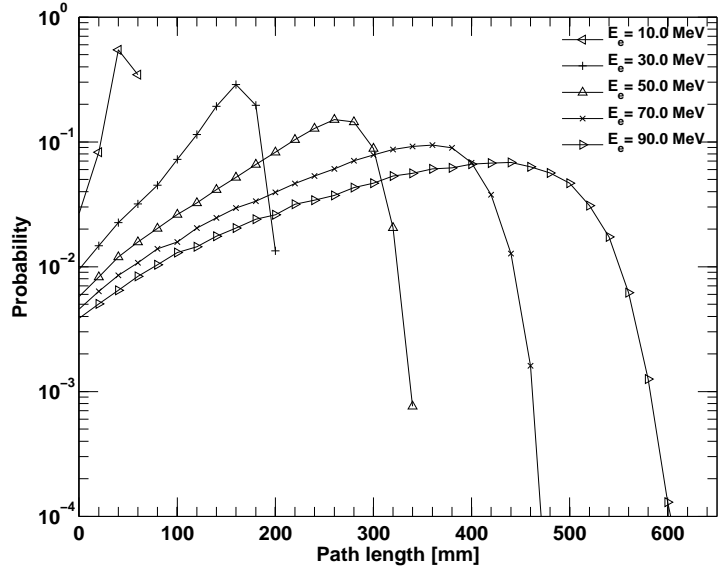}
\caption{Probability density function of the path length of positrons with different kinetic energies.}
\label{fig:ranges}
\end{center}
\end{figure}

\begin{figure}[h!]
\begin{center}
\includegraphics[width = 0.7\textwidth]{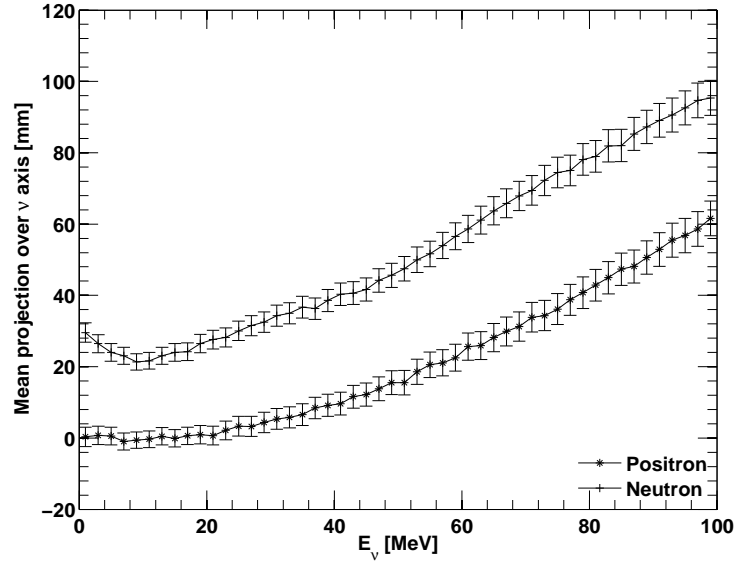}
\caption{Mean positron and neutron paths projections along the $\nuebar$ direction as a function of the neutrino energy.}
\label{fig:positron_neutron_projection}
\end{center}
\end{figure}

Naturally the mean path length increases with the energy, with a peak value ranging from a few tens of mm to a several hundreds of mm at higher energies. Figure~\ref{fig:positron_neutron_projection} shows our results for the mean positron path length projected onto the incoming neutrino direction as a function of the neutrino energy. It ranges from a few mm for $E_\nu<$15~MeV up to about 40~mm at 80 MeV. Note that for low energy neutrinos ($E_{\nu}<$10~MeV), the projected path length is $\sim 0.05$~mm, in agreement with the expected value for reactor neutrinos~\citep{Apollonio:1999jg}.\\

\begin{figure}[h!]
\begin{center}
\includegraphics[width = 0.5\textwidth]{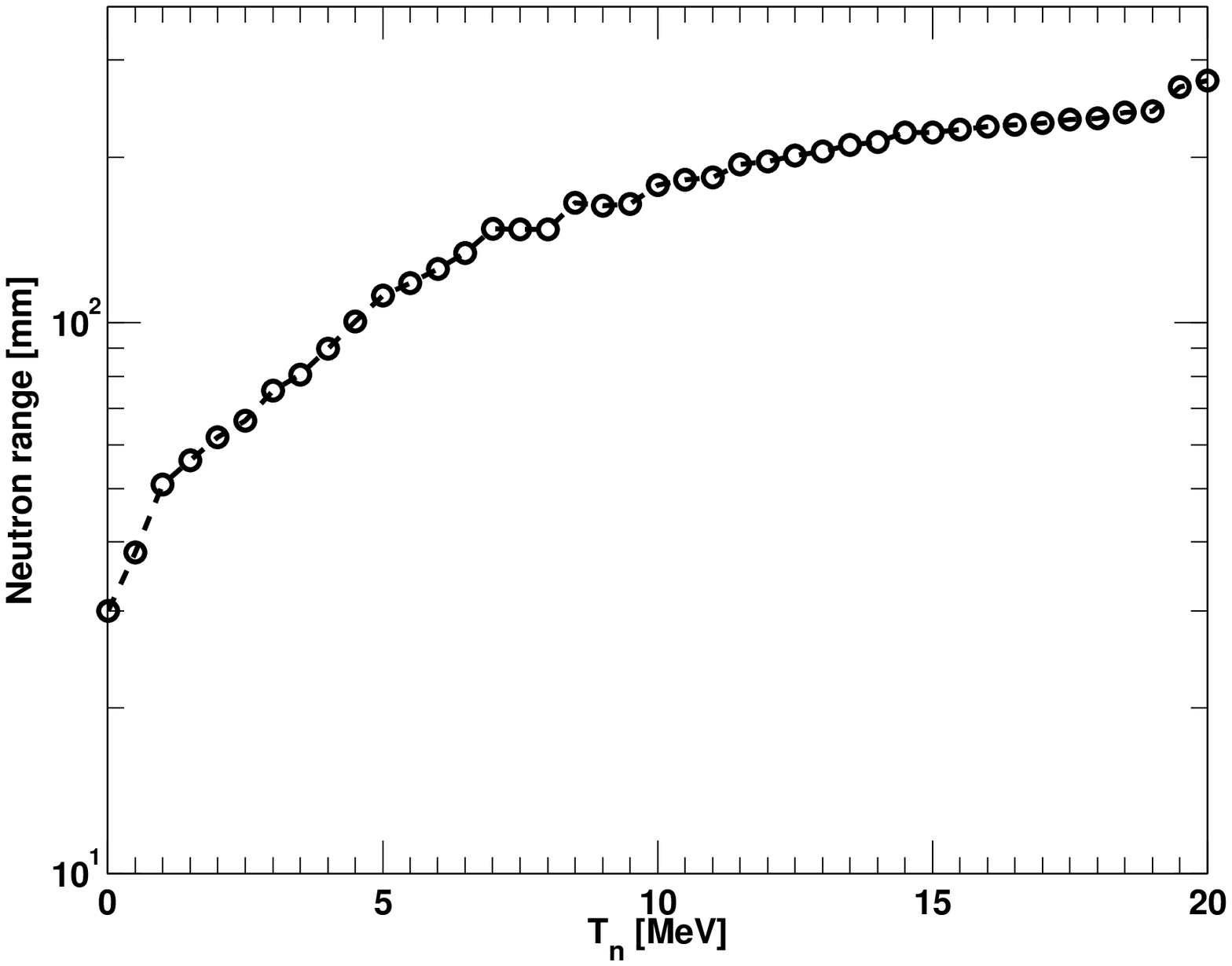}
\includegraphics[width = 0.475\textwidth]{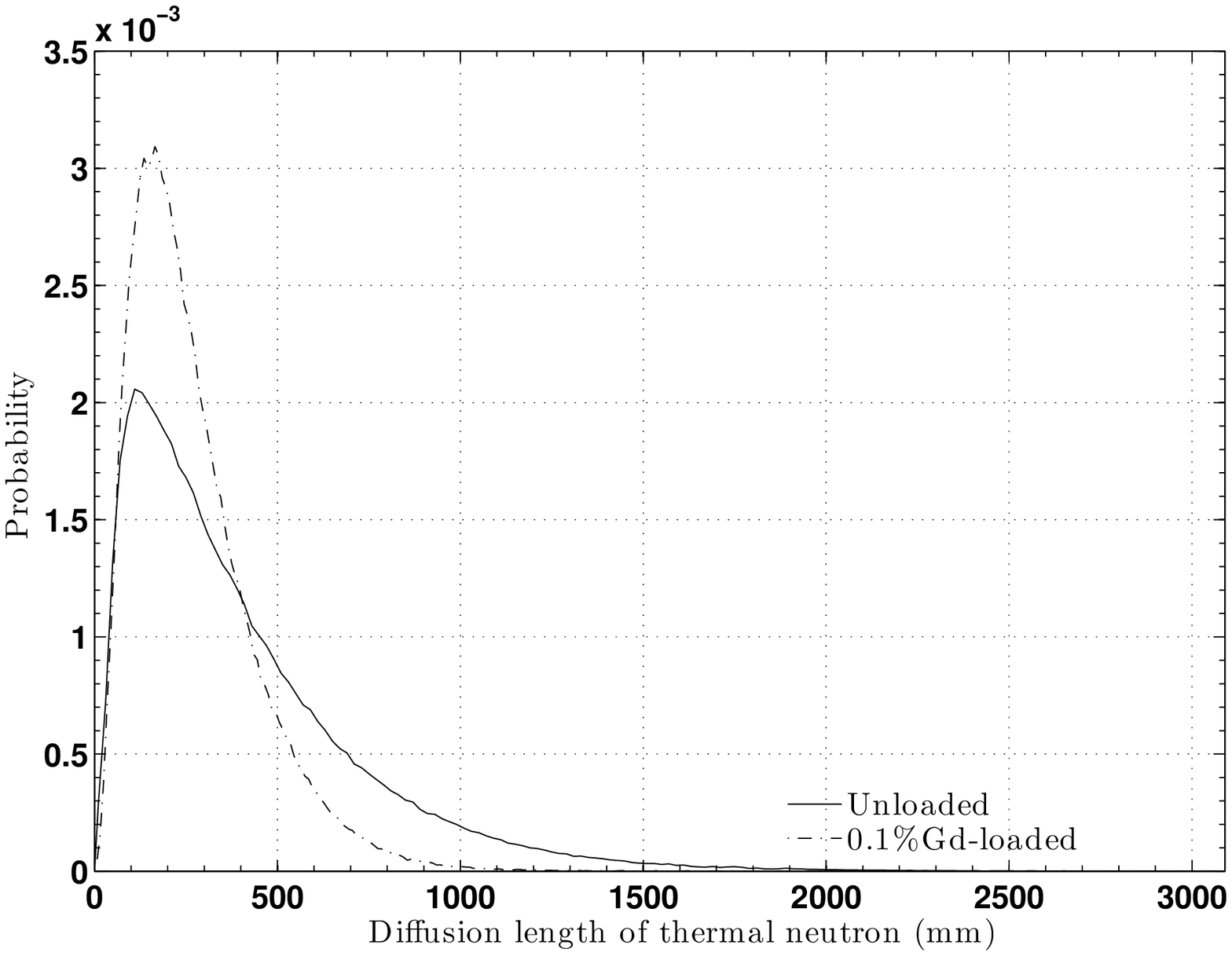} 
\caption{Left: Mean neutron path length before diffusion as a function of the neutron kinetic energy T$_{\text{n}}$. Right: Diffusion length of thermalized neutrons in liquid scintillator with and without Gd-doping.}
\label{fig:capture_n}
\end{center}
\end{figure}

The neutron, emitted simultaneously with the positron, loses its energy through elastic scattering off light nuclei of the scintillator medium until it thermalizes with the medium. This non-isotropic process, depending both on the neutron energy and on the 
liquid composition, preserves a memory of the initial neutrino direction. The mean path length between the neutron creation and its thermalization obtained from our simulation is displayed in Figure~\ref{fig:capture_n} (left), for different neutron kinetic energies.
After being moderated the neutron scatters at thermal energies until it gets captured. The path length of this diffusion process can be seen in Figure~\ref{fig:capture_n} (right).

The diffusion is an isotropic random walk spreading the distribution of the neutron capture point.
Our simulations have shown that the variation in path length from a non-doped scintillator liquid compared to another is negligible. Figure~\ref{fig:positron_neutron_projection} shows our results for the average neutron path length projected onto the incoming neutrino direction as a function of the neutrino energy. It ranges from 25 mm for $E_\nu<$15~MeV up to about 80~mm at 80 MeV. It is worth noting that we recover the results commonly used for reactor neutrino experiment for which the projection is expected to be $\sim$20 mm for neutrino energies up to 10 MeV.
Most of the neutron captures occur on an hydrogen atom thus emitting a 2.2~MeV gamma ray whose energy deposition constitutes the delayed event. Since this gamma is emitted isotropically and its mean path length in the liquid is about 20~cm, it smears the capture vertex location.\\

\subsection{Direction retrieval}

\label{subsec:IBD_directionality}

\subsubsection*{Method}
\label{subsubsec:fit}

\begin{figure}[h!]
\begin{center}
\includegraphics[width = 0.5\textwidth]{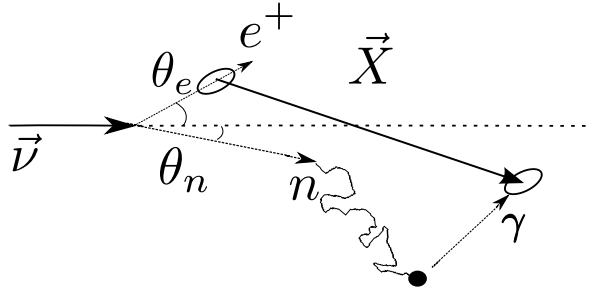} 
\caption{Sketch of the Delayed-Prompt vector $\vec{X}$ along the direction of the incoming neutrino.}
\label{fig:prdel_vect_fig}
\end{center}
\end{figure}

The information returned by our simulation is only the visible energy and the reconstructed vertices of the prompt and delayed events, likewise a real supernova observation in a LLSD. From these vertices, we obtain the Delayed-Prompt unit vector $\vec{X}$ for each neutrino event having its origin at the positron reconstructed position and pointing to the neutron capture reconstructed location, as displayed in Figure~\ref{fig:prdel_vect_fig}. 

\begin{figure}[h!]
\begin{center}
\center \includegraphics[width = 0.7\textwidth]{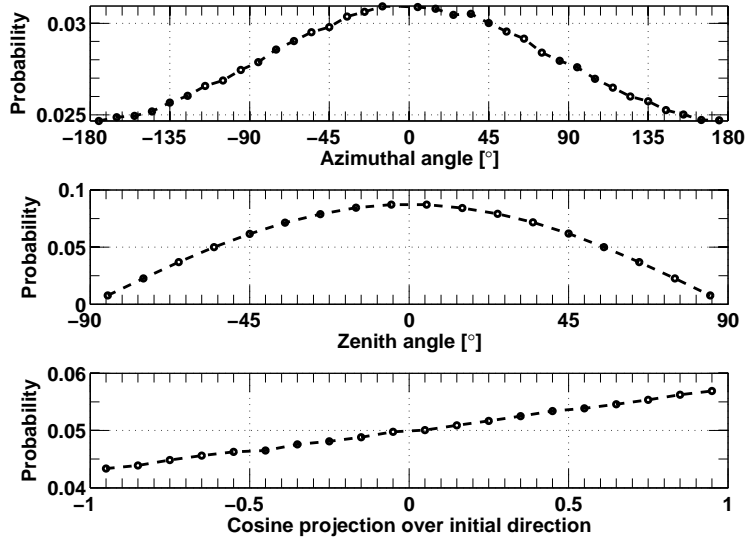}
\caption{Probability density functions of azimuthal angle (top), zenith angle (center) and cosine projection on the neutrino direction (bottom) distributions extracted for a supernova pointing towards the (0,0) spherical coordinates in the detector frame. The angular distributions are centered on the (0,0) spherical coordinates. Neutrino energies are drawn from a GVKM spectrum.}
\label{fig:3angular-dist-GVKM}
\end{center}
\end{figure}

We then compute a direction $\left( \phi,\theta \right)$ for each event with $\phi$ and $\theta$ respectively the azimuthal and zenith angles in the detector reference frame. The angular distributions extracted from a supernova simulation are displayed in Figure~\ref{fig:3angular-dist-GVKM}. These distributions are respectively the azimuthal angle distribution, the zenith angle distribution and the cosine projection on the neutrino axis expressed as:
\begin{equation}
  \cos{\delta} = \frac{\vec{X} \cdot \vec{\nu}}{\vert \vert \vec{X} \vert \vert \times \vert \vert \vec{\nu} \vert \vert},
\label{eqn:cosine_proj}
\end{equation}
with $\cos{\delta}$ the normalized scalar product of $\vec{X}$ the Delayed-Prompt vector and $\vec{\nu}$ the neutrino direction.
All simulations were carried out using a conservative vertex resolution of 20~cm. Tests with a 15~cm vertex resolution were performed without yielding significant improvements.

In order to consider all the relevant features of the IBD kinematics we reconstruct the direction with a $\chi^{2}$ minimization method comparing a model and a supernova dataset. Both are simulations of supernova neutrinos, performed with our toy Monte Carlo described in Appendix~\ref{section:toyMC}, with a user-defined energy spectrum and cross section. The model dataset is computed with a large number of events in order to smoothen statistical effects and generated along an arbitrary direction.
On the other hand, the supernova dataset has a realistic statistic corresponding to the number of IBD interactions in a given detector and is generated along a user-defined direction $\left( \phi_{0},\theta_{0} \right)$. By rotating the model distribution across the sky, our algorithm finds the best-fit angles $\left( \phi,\theta \right)$ corresponding the minimum value of the $\chi^{2}$ expressed as:\\

\begin{equation}
\chi^{2} \left( \phi,\theta \right) = \sum_{i}^{N} \sum_{j}^{N} \left( Y_{i} - M_{i}\left( \phi,\theta \right) \right) \left(V^{-1}\right)_{ij} \left( Y_{j} - M_{j}\left( \phi,\theta \right) \right),
\label{eqn:chi2_equation}
\end{equation}
where $Y_{i}$ and $M_{i}\left( \phi,\theta \right)$ are the concatenation over N bins of the three angular distributions (azimuthal, zenith and cosine projection) of the supernova and model datasets, respectively. $V_{ij}$ is the N$\times$N covariance matrix made of two independent blocks accounting for the bin-to-bin statistical uncertainty of the azimuthal and zenith angle distribution and a third block, correlated to the first two, accounting for the bin-to-bin statistical and systematic uncertainties of the cosine projection distribution expressed in Eq.~\ref{eqn:cosine_proj}. The $V_{ij}$ covariance matrix has been generated using 2,000 independent datasets of 5,000 neutrino events each, then normalized to one event. The resulting unitary covariance matrix is then normalized to the number of events in the supernova dataset being used in each $\chi^{2}$ minimization.

The best-fit angles $\left( \phi,\theta \right)$ corresponding to a minimum of $\chi^{2}$ are then stored and the process is repeated with a new supernova dataset having the same statistic. From all these trials, we obtain distributions of reconstructed azimuthal and zenith angles whose means and variances correspond to the reconstructed supernova directions and associated angular errors.

\subsubsection*{Results with a single detector}
\label{subsubsec:direction_single_detector}

\begin{figure}
  \centering
  \subfigure[]{\includegraphics[width = 0.44\textwidth]{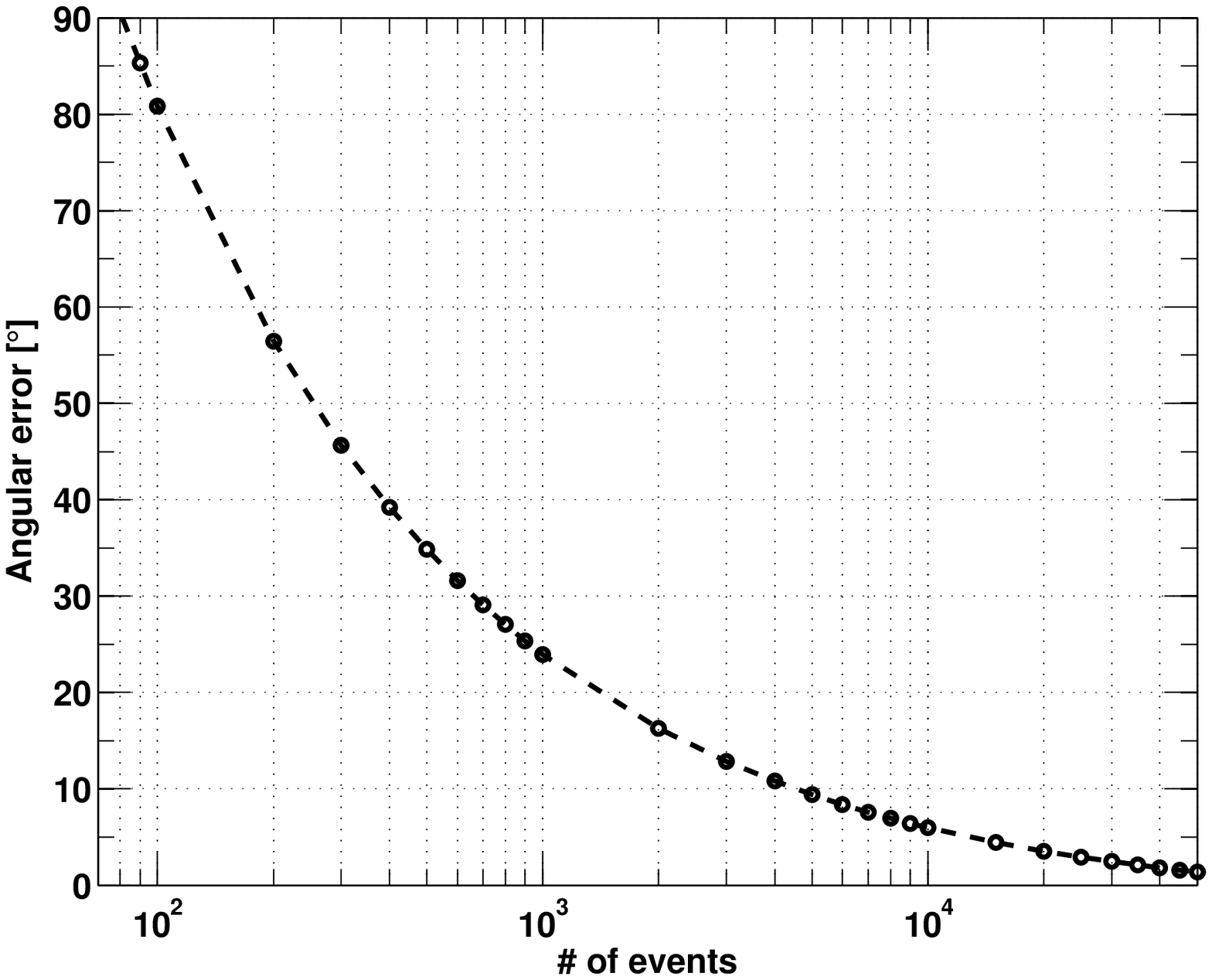}}
  \subfigure[]{\includegraphics[width = 0.55\textwidth]{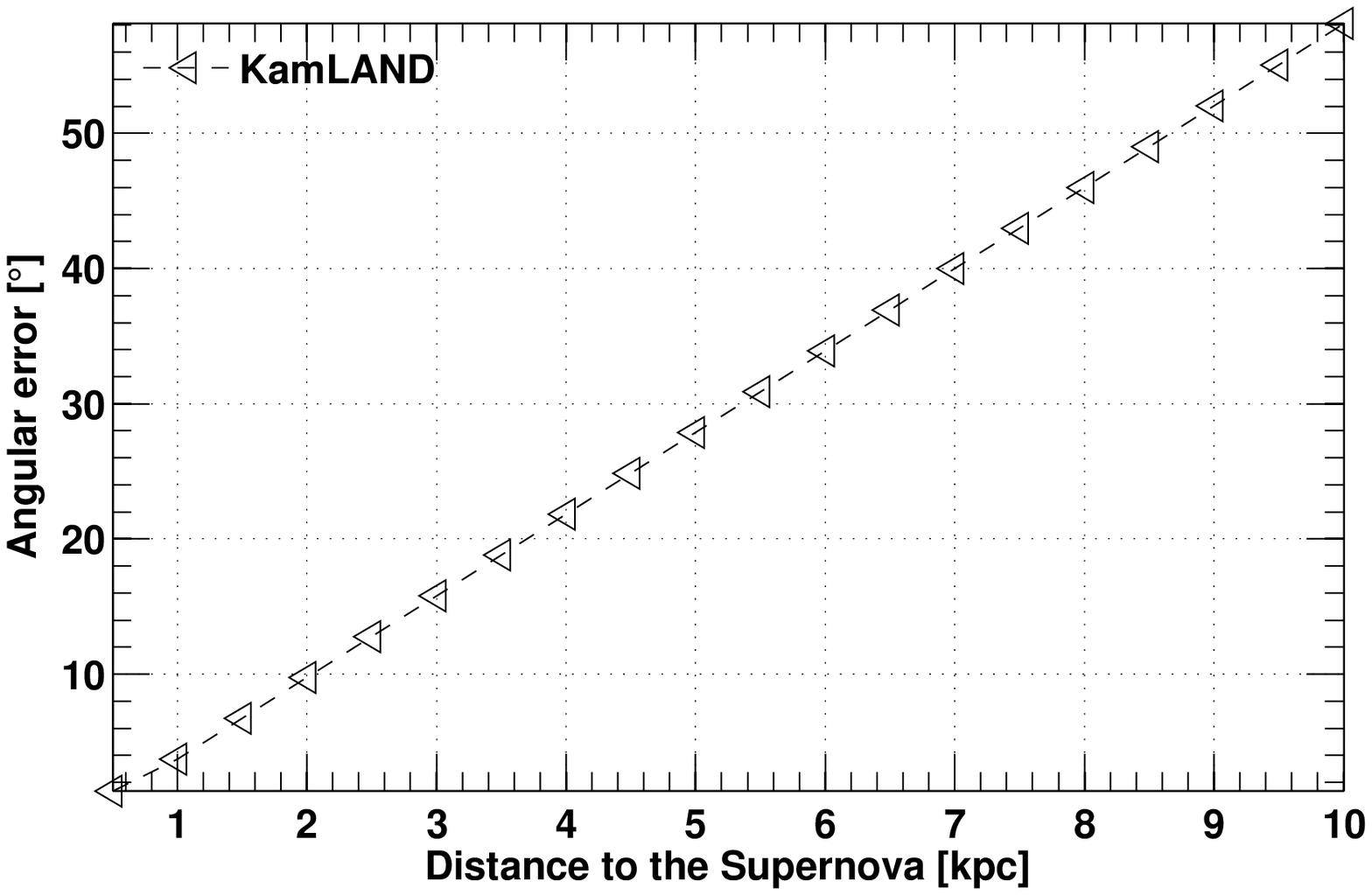}}
  \caption{a: Angular uncertainty as a function of the number of detected events. b: Angular uncertainty as a function of the distance in kpc in the KamLAND detector.}
 \label{fig:angular_error_wrt_distance_nbevt}
\end{figure}

We now apply the method previously described to a single detector. We considered KamLAND in this first application. We generated several datasets with different number of events and computed their associated angular uncertainty using the method previously described.
Figure~\ref{fig:angular_error_wrt_distance_nbevt} shows the interpolation of our results, best fitted by an inverse square root function, and describes the behavior of the angular uncertainty with respect to the number of detected events. Using this curve and the expected number of events computed from Section~\ref{sec:LLSD_detection}, we associate an angular uncertainty on the supernova reconstructed direction for KamLAND at different distances as shown in Figure~\ref{fig:angular_error_wrt_distance_nbevt}. Several simulations were carried out using the different flux shapes presented in Section~\ref{sec:SN_neutrino_signature} and no sensible differences in the results have been found.
With a number of events leading to an angular uncertainty greater than 90$^{\circ}$, we consider that no directional information can be extracted and consequently discard the concerned detector from our directionality reconstruction. On the other hand, the detection of more than 50,000 events would lead to a trigger rate too high to be withstood, hence we arbitrarily consider it as our detection limit. Therefore, any observation of a higher neutrino rate in a single detector would, in our analyses, not bring additional direction information and the reconstruction error could not reach lower values. Between these two limits, the angular uncertainty increases with the distance and with the inverse of the number of events. The statistical nature of the IBD directionality process is well shown since several thousands of events are required to obtain a decent reconstruction uncertainty. As a consequence, the pointing potential of a single detector such as KamLAND gets degraded as soon as the disctance exceeds a few kpc, hence the interest of combining several detectors to improve the accuracy.

\subsubsection*{Combining large liquid scintillator detector data}
\label{subsubsec:direction_globalfit}

Higher sensitivities on supernova position reconstruction are reachable adding the directional information provided by several detectors. Each of the detectors previously described in Section~\ref{sec:LLSD_detection} could detect a supernova assuming it leads to a sufficient amount of detected events. The addition of these signals increases the supernova detection potential in two ways. 

First, observing a coincident supernova signal in several detectors across the globe strengthens the confidence that a supernova has occurred. This is the current role of the SNEWS network that will be discussed in Section~\ref{sec:SNEWS_network}. Furthermore the time difference between the supernova signals coming from those detectors could give us a first hint of localization using the triangulation method as discussed in~\citep{Beacom:1999}. This method might not be accurate enough at typical supernova distances since it requires a good understanding of the luminosity time distribution and is strongly statistic dependent. A significant work of time calibrations between detectors would also be mandatory.

Second, one can combine the reconstructed angular directions and errors provided by those detectors to improve the localization of the detected supernova.

\begin{table}[h!]
\begin{center}
\caption{List of individual detectors and their corresponding datasets.}
\footnotesize
\begin{tabular}{ c | c | c | c | c }
\hline
 & Existing & SNEWS & Near future  & All \\
\hline
LVD			&		& 		& 		& 	 \\
KamLAND			&$\times$	&$\times$	&$\times$	&$\times$\\
SNO+			&$\times$	&$\times$	&$\times$	&$\times$\\
Borexino		&$\times$	&$\times$	&$\times$	&$\times$\\
Daya Bay		&$\times$	&$\times$	&$\times$	&$\times$\\
Double Chooz		&$\times$	&		&$\times$	&$\times$\\
RENO			&$\times$	&		&$\times$	&$\times$\\
NOvA (far)		&		&		&		&	\\
NOvA (near)		&		&		&		&	\\
MiniBooNE		&		&		&$\times$	&$\times$\\
JUNO			&		&		&$\times$	&$\times$\\
LENA			&		&		&		&$\times$\\
\end{tabular} 
\label{table:detectors_sets}
\end{center}
\end{table}

In the context of a global analysis with N detectors combined, all the individual angular ``cones'' of 1$\sigma$ azimuthal and zenith uncertainties must be combined into a single one. To do so, we combined the individual angular errors and expressed the average angular uncertainty as:
\begin{equation}
   \bar{\sigma} = \left(\sum\limits_{i=1}^{N}{\frac{1}{\sigma_{i}^{2}}}\right)^{-\frac{1}{2}},
\end{equation}
with $\sigma_{i}$ the 1$\sigma$ uncertainty on the supernova direction reconstructed in the i$^{\text{th}}$ considered detector.

\begin{figure}[ht!]
\begin{center}
\includegraphics[width = 0.95\textwidth]{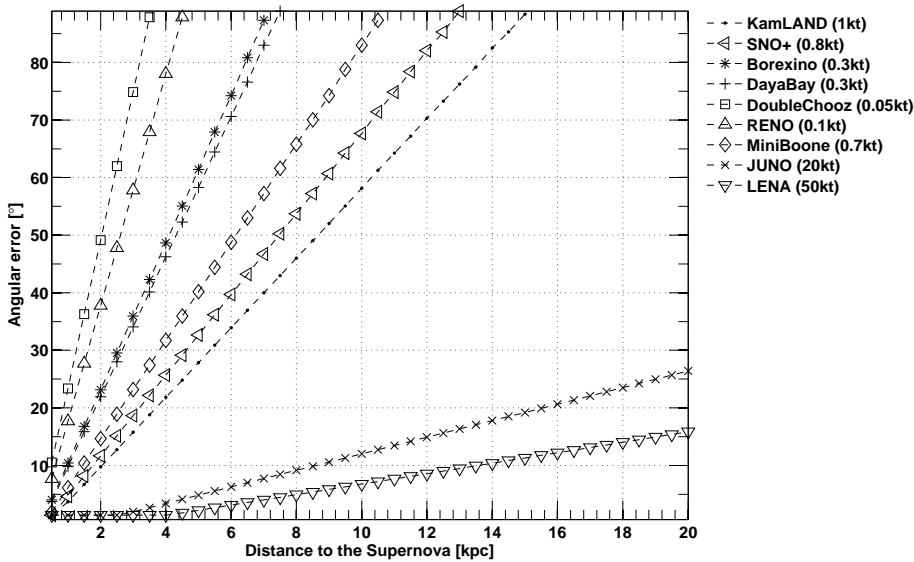} 
\caption{Angular uncertainty as a function of the supernova distance for different detectors and their associated masses.}
\label{fig:error_dist_all_det_indiv}
\end{center}
\end{figure}

\begin{figure}[ht!]
\begin{center}
\includegraphics[width=.95\textwidth]{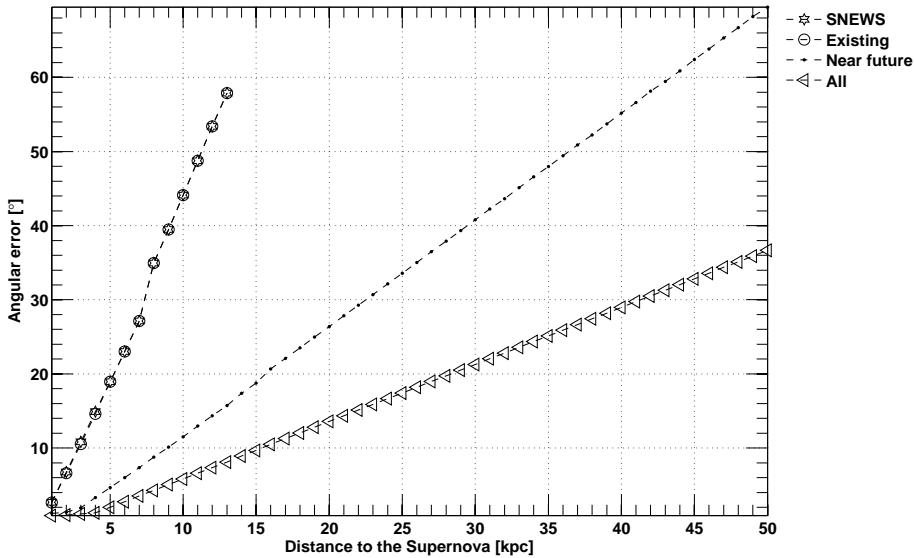} 
\caption{Angular uncertainty as a function of the supernova distance for different detector combinations. The mean reconstruction error for combined detectors is computed using the combination of the angular reconstruction error associated with each individual detector. Due to the small event rate expected in Double Chooz and RENO, the ``SNEWS'' and ``Existing'' datasets yield very similar results.}
\label{fig:error_dist_all_det_sets}
\end{center}
\end{figure}

In the following analysis, we considered 5 sets of combined detectors, summed up in Table~\ref{table:detectors_sets}. The ``Existing'' dataset is the combination of all current LLSD, ``SNEWS'' consists of the four LLSD belonging to the SNEWS network as detailed in Section~\ref{sec:SNEWS_network}, ``Near future'' is the sum of all current LLSD with the addition of MiniBooNE and JUNO, planned to operate within the next 5 years and finally ``All'' consists of the sum of all detectors, current and future.
Results are displayed in Figure~\ref{fig:error_dist_all_det_indiv} for individual detectors and Figure~\ref{fig:error_dist_all_det_sets} for the sets of combined detectors.

Considering a supernova located at 10~kpc, no current detector could provide a valuable direction information other than the hemisphere in which the supernova occurred. However, when combining several of these detectors, this same supernova direction is likely to be reconstructed within a 45$^{\circ}$(68\% C.L.) cone. While this angular opening is still larger than most of the optical instruments can accommodate, the addition of JUNO in a near future would reduce the uncertainty to 12$^{\circ}$(68\% C.L.), hence reaching a relevant level, almost competitive with the current Super-Kamiokande detector.
With the conservative assumption that any galactic gravitational stellar collapse will occur within a 20~kpc radius, the addition of current and future detectors could reconstruct its position within a 14$^{\circ}$(68\% C.L.) cone.

\section{Improving the SNEWS network?}
\label{sec:SNEWS_network}

The coincident observation of a burst in several neutrino detectors would be a robust early warning of a forthcoming visible galactic core-collapse supernova. The main goal of the SuperNova Early Warning System (SNEWS) is to provide the astronomical community with this early alert of a supernova~\citep{Antonioli:2004zb}.
Each experiment belonging to SNEWS has its own supernova real-time trigger system dedicated to detect a burst of events that might be linked to a supernova. When the trigger requirements are met, an alert is sent to the global SNEWS server and is compared with possible alerts coming from other experiments. If a coincidence is found within a time window of 10~seconds, a global alert is sent to the SNEWS mailing list, universally accessible via snews.bnl.gov. In 2014, Daya Bay joined the SNEWS network already composed of Super-Kamiokande, LVD, Borexino, SNO+, KamLAND and IceCube. Adding Double Chooz, RENO and NO$\nu$A to the network would increase the number of detected supernova neutrinos, thus strengthening the confidence level of an hypothetical alert.

The SNEWS network is set up to provide information that a supernova has been detected via its neutrinos to astronomers. Although this information is mostly characterized by the observation of a burst of events in coincidence in several experiments, a directional information can be sent as well, if provided. Without this information, the supernova location remains unknown until its observation. Knowing this position, even roughly, over the sky would give astronomers a consequent head start and could lead to the observation of a complete supernova process, including luminosity rise. 

As shown in Section~\ref{subsubsec:direction_globalfit}, even though current LLSDs in SNEWS can provide directional information for nearby supernovae, Super-Kamiokande remains the best detector to locate them. However, with the replacement of SNO heavy water with liquid scintillator and before the construction of Hyper-Kamiokande~\citep{Abe:2011ts}, it stands alone as the world's only large \v{C}erenkov detector. Relying solely on Super-Kamiokande might not be enough to ensure a permanent watch on supernovae. Any maintenance operation or reconstruction software issue could put astronomers in the blind in case of a long-awaited supernova detection.

With the help of SNEWS's current and future liquid scintillator detectors, this scenario is unlikely to happen as the directional information of several detectors that observed the supernova can be combined. In the case of all SNEWS detectors operating at once, the Super-Kamiokande direction reconstruction will be improved by the combined reconstruction of all LLSDs. With only 4 LLSDs (KamLAND, Borexino, SNO+ and Daya Bay) capable of direction reconstruction, SNEWS is currently able to locate a 8~kpc supernova within a barely practical 35$^{\circ}$(68\% C.L.) wide cone. However, with the addition of all current LLSDs presented in Section~\ref{subsec:running_detectors} as well as JUNO and MiniBooNE in a near future, the opening of this cone could be reduced to 9$^{\circ}$(68\% C.L.) thus reducing by a factor 15 the area of the sky region of interest.

In order to join the SNEWS network, each experiment would have to meet a few requirements, one of the most important being the robustness of its data acquisition system (DAQ), i.e. its capability to quickly and efficiently tag a supernova burst. In order to send a reliable alert to the SNEWS global server, each detector's DAQ needs to be modified to operate on real-time mode and detect any unusual trigger rate increase. In the case of NO$\nu$A, the detector's trigger is directly synchronized with the NuMI beam trigger hence the necessity of operating the detectors in a ``free'' mode without any external trigger as presented in Ref.~\citep{Ayres:2004js}. An example of the implementation of such a trigger for the Daya Bay detectors is detailed in Ref.~\citep{HanyuWeifortheDayaBay:2013oma}.

Implementing a combined pointing ability within the network might be more challenging. While the detection of a supernova burst solely relies on the observation of a burst of events in several detectors in coincidence, its pointing towards the sky requires energy and position reconstructions of these events. Most detectors operate on a run-by-run basis and usually analyze large datasets after long periods of data taking. Providing a neutrino direction within a time scale of a few hours would require offline and real time data analysis. This might be achieved by installing separate and independent online data acquisition and analysis chains capable of supporting large data rates and fast energy and position reconstruction based on the current analysis.


\section{Retrieving astrophysical information from supernova neutrinos}
\label{sec:energy_fits}
Beyond our main goal of determination of the supernova direction, we discuss in this section additional information that could be retrieved by LLSDs, using the IDB process. In particular, we consider the determination of the supernova distance and then the assessment of the neutron star mass. The latter was investigated in Ref.~\citep{Sato:1987yi} in the case of SN1987A. 
Clearly it would also be of interest to determine how well one could identify the average energy and pinching parameters of the supernova neutrino fluxes associated with the different phases of the explosion (neutronization burst, accretion and cooling). This has been studied e.g. in~\cite{Vaananen:2011bf} by combining one- and two-neutron emission channels in lead-based detectors. However a similar study for scintillator detectors goes beyond the scope of the present manuscript. Just as an example, we show how well one could  discriminate the two supernova neutrino fluxes used in the present study.

\subsection{Measurement of the supernova distance}
\label{subsec:distance_fit}
Since, for a given supernova, the number of detected events varies with the inverse of the squared distance, the determination of its distance sounds straightforward in principle. To better understand the impact of statistical and systematic uncertainties, we developed a Monte-Carlo simulation to generate and fit detected energy spectra with a $\chi^{2}$ minimization method. This allows us to compare fake data fitted with the GVKM flux under various hypotheses. The main systematics considered are uncertainties on the neutron star features. Indeed the number of emitted neutrinos depends on the difference between the progenitor star and the neutron star gravitational binding energies. However, the first term can be considered negligible and the neutrino fluence therefore depends on the neutron star binding energy\footnote{In Sections~\ref{subsec:distance_fit} and~\ref{subsec:NS_mass_fit} an approximated relation of the gravitational binding energy of a sphere of uniform density is employed. We use it here for illustrative purposes, since the determination of the supernova distance and of the neutron star mass are not the main purposes of the present work.} $E_{B}~=~\frac{3}{5}\frac{M_{ns}^{2}}{R_{ns}}$ with $M_{ns}$ and $R_{ns}$, being the mass and radius of the neutron star~\citep{Zuber:2003}. In the literature these parameters are usually set to $M_{ns} = 1.4~M_{\odot}$ and $R_{ns} = 10$ km~\citep{Giunti:2007ry}. From the average observed neutron star masses and radii computed in Ref.~\citep{Lattimer:2004sa} and Ref.~\citep{Guillot:2013wu}, we conservatively consider a relative uncorrelated systematic uncertainty of 10\% for both parameters. We simulated GVKM-based supernovae bursts located from 1 to 10~kpc away and reconstructed their distances and associated uncertainties (stat. + syst.). Hereafter, a conservative energy resolution of 6\% in the whole energy range is taken into account. The distance is reconstructed without bias and with an uncertainty of 40\% and 15\% at 10~kpc for a 1~kt and a 20~kt detector, respectively. These results are valid for the spectral shapes of the GVKM and Livermore fluxes presented in Section~\ref{sec:SN_neutrino_signature}.

\subsection{Measurement of the neutron star mass}
\label{subsec:NS_mass_fit}
A large set of neutron star masses have been previously measured and are listed in Ref.~\citep{Lattimer:2004sa}. These measurements provide relative uncertainties at the 0.1\% level on the neutron star mass for binary systems. However, an isolated system such as a single supernova without companion would not provide as much precision. Hence, our interest to retrieve the neutron star mass from the number of detected neutrino events.
We now fix the supernova distance, assumed to be known by another conventional technique and try to retrieve the neutron star mass from the number of detected events and the expression of the neutron star gravitational energy. 
In a practical case, the measurement of the distance could be achieved through the Expanding Photosphere Method~\citep{Schmidt:1992yr}. This technique led to a 10~\% uncertainty of the SN1987A distance. Using this as a reference, we fix the uncertainty on the reconstructed distance to 10\%. We also fix the neutron star radius and associated uncertainty, as in Section~\ref{subsec:distance_fit}. 
We can thus assess how well one could measure the neutron star mass by fitting the energy distribution of incoming neutrino events as in Section~\ref{subsec:distance_fit}. 
We generated neutrino supernova spectra with statistics corresponding to different masses between 1 and 2 solar masses. For a canonical supernova located at 10~kpc, the mass could be measured with an uncertainty of 16\% and 13\% for a 1~kt and a 20~kt scintillator detector respectively.

\subsection{An example: Disentangling different supernova flux shapes}
\label{subsec:model_discri}
\begin{figure}[h!]
\begin{center}
\includegraphics[width = 0.49\textwidth]{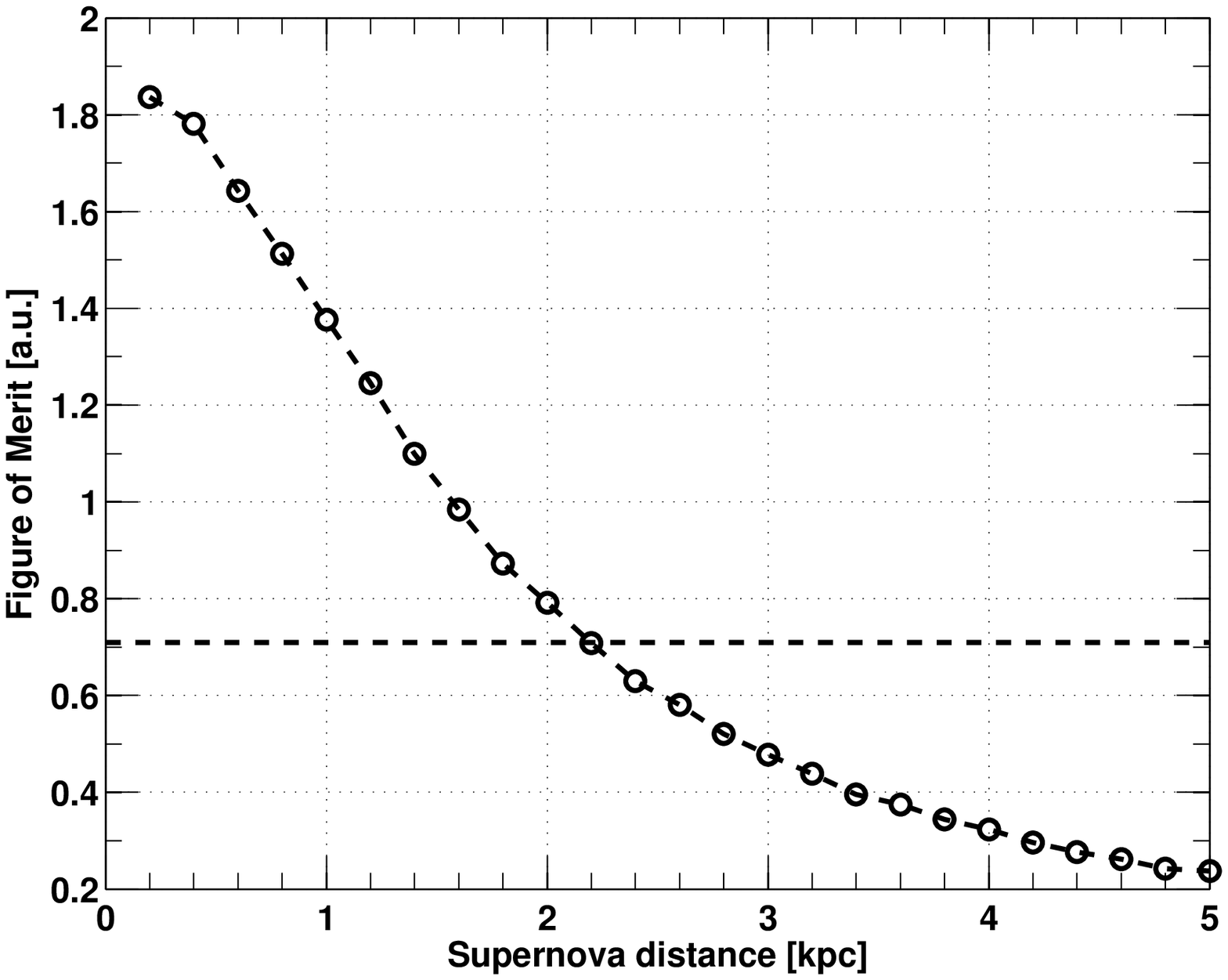}
\includegraphics[width = 0.49\textwidth]{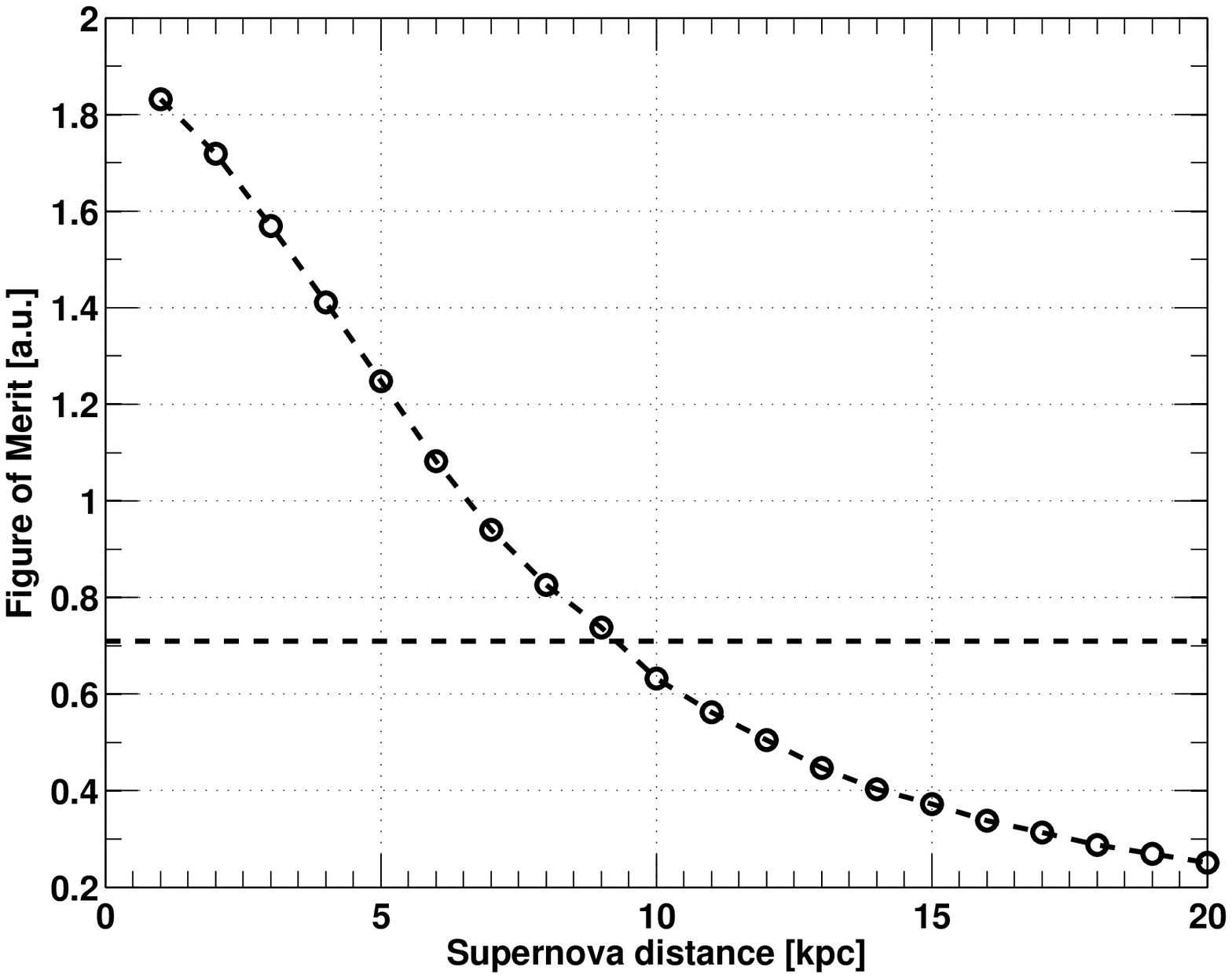}
\caption{Figure of merit for disentangling GVKM and Livermore energy distributions as a function of the supernova distance for a 1~kt (left) and a 20~kt (right) detector. The dotted line represents the upper limit over which a GVKM spectrum is reconstructed with at least 95\%~C.L.}
\label{fig:FoM_liv_gvkm}
\end{center}
\end{figure}

The far location of SN1987A in the Large Magellanic Cloud has provided information on the total gravitational energy radiated as neutrinos and on the neutrino temperatures, consistent with predictions (see Ref.~\citep{Vissani:2014doa} and references herein). However, the statistic was not large enough to precisely determine the supernova neutrino energy spectrum. In what follows we study the possibility to discriminate between two flux shapes, GVKM and Livermore, with the current and future generation of LLSDs. 
We statistically compare spectra generated from a GVKM distribution with both GVKM and Livermore fluxes. To simplify the problem as well as remain conservative, we normalize both fluxes to the same fluence, only their spectral shapes differ. We then compare the $\chi^{2}$ distributions of 5,000 simulated experiments fitted with the GVKM and Livermore fluxes. The two $\chi^{2}$ distributions are generated for supernova distances ranging from 0 to 10~kpc. The separation power between the two is expressed through the following figure of merit: $\text{F.o.M.} = \frac{\mu_{2} - \mu_{1}}{2.35 \times \left( \sigma_{2} + \sigma_{1} \right)}$ with $\mu_{1,2}$ and $\sigma_{1,2}$ the respective mean and standard deviation of the two distributions. This figure of merit has been carefully chosen for its efficiency in disentangling Gaussian-like probability distributions. It has been used in the Nucifer experiment to separate neutron and gamma backgrounds via the Pulse Shape Discrimination (PSD) technique. More details can be found in Ref.~\citep{Gaffiot:2012lja}.
Figure~\ref{fig:FoM_liv_gvkm} shows the behavior of this separation power parameter as a function of the increasing distance for a 1~kt and a 20~kt detector. The reconstruction of a GVKM spectrum at 95\%~C.L. can be achieved up to a distance of 2.2 kpc for a 1~kt detector such as KamLAND and 9.5~kpc for a 20~kt detector such as JUNO.

\section{Discussion}
\label{sec:discussion_outlook}

In this paper, we discussed the interest of combining worldwide large liquid scintillator detectors  to efficiently reconstruct a supernova location. Such an information would be precious to the SNEWS network in order to provide not only an alert that the optical observation of a supernova is imminent but also that it is likely to occur in a restricted region of the sky. While for a typical supernova located at 10~kpc, a 45$^{\circ}$(68\% C.L.) reconstruction cone would be currently too large to be entirely enclosed within the aperture of an optical telescope\footnote{Note that the Moon's angular diameter over the sky is 0.5$^{\circ}$ hence enclosed in a 0.25$^{\circ}$ radius cone.}, it would focus astronomers attention to a particular field of view, enhancing the probability for a comprehensive optical observation. The impending Zwicky Transient Facility (ZTF)~\cite{Bellm:2014} will be able to completely scan this region of the sky within two hours. The presence of several telescopes based on the design described in Ref.~\cite{Adams:2013ana} would provide a high probability of successful observation as well. The implementation of future detectors would soon reduce the region of interest to a 12$^{\circ}$(68\% C.L.) radius cone. In the case of a nearby supernova 2~kpc away, a single detector such as KamLAND could point its direction within a 10$^{\circ}$(68\% C.L.) cone and the combinations of current and/or future detectors could reduce this uncertainty below 5$^{\circ}$(68\% C.L.).

Besides providing a relevant direction reconstruction, a wealth of information can be obtained through IBD events. In particular, in Section~\ref{sec:energy_fits} we have discussed the extraction of information on the supernova distance and the mass of the neutron star. We also discussed how accurately two  flux models could be discriminated, as a function of the supernova distance.

Though it has hardly been discussed in this paper, the observation of a large quantity of neutrinos in several large detectors will help improve the direction reconstruction via triangulation technique. The combined data from several time-calibrated detectors scattered over the globe would be contained within the 40~ms time window corresponding to the Earth's crossing. The detection of a full neutrino spectrum at each moment of the supernova luminosity rise and decay would decrease any statistical uncertainty related to the luminosity rise time thus making triangulation a viable technique as explained in Ref.~\citep{Beacom:1999}. 
Since only IceCube, Super-Kamiokande and NO$\nu$A's far detector are currently operating with active masses relevant for triangulation, it might not be a suitable method as we speak. However, with the impending construction of large detectors such as JUNO, LENA or Hyper-Kamiokande, it could be considered as reliable as the other methods.
Another pointing method relying on the observation of spectral distortions caused by matter effects in Earth is discussed in Ref.~\citep{Scholberg:2009jr}.

Recent technological developments have been successfully carried out in the field of detection materials. For example, detectors using liquid scintillators mixed and dissolved into water are currently being designed. A detailed study of such technology and its potential physics applications can be found in Ref.~\citep{Alonso:2014fwf}. Besides combining the technical advantages of both materials, a water-based liquid scintillator detector also provides two directional information. Indeed, enhanced by the possible use of fast photodetectors such as LAPPD's (Large Array Picosecond PhotoDetectors), the collection of \v{C}erenkov and scintillation light produced by the both IBD and electron scattering processes would exploit the full directional information carried by neutrinos.

We would like to draw attention to technical issues that might arise along with the high event rates in large detectors. Underground detectors are designed to operate in an ultra low background environment with typical trigger rates of a few hundred Hz. The detection of tens of thousands of neutrinos even distributed in time over a 10~s window with a strong peak at 1~s may lead to a non-negligible dead-time or, worst case scenario, to a DAQ crash. On the other hand, the NO$\nu$A far detector, since it operates at the surface, is subjected to a 300~kHz background rate from cosmic rays at 10~MeV. While thanks to its fine-grained segmentation, it could separate IBD events from cosmic muons tracks, an efficient supernova tagging would place significant demands on the DAQ in terms of dead-time and data readout and analysis rates. In order for any of these detectors to efficiently collect a large number of neutrinos in a short period of time and increase its pointing ability, considerable efforts will be required to improve the current state of data acquisition.

\section{Conclusion}
\label{sec:conclusion}

In conclusion simultaneous detection of several hundreds or thousands of neutrinos via several processes will bring an enormous amount of information about stellar evolution and neutrino flavor conversion occurring in extreme media. The directional information extracted from these same neutrinos is a precious input for astronomers and could lead to a prompt and therefore comprehensive observation of a galactic supernova in several domains of the electromagnetic spectrum. To link these two observations and recover as much data as possible from the next and long-awaited supernova in our galaxy, the SNEWS network performs a constant watch that could be improved by incorporating all large neutrino detectors, all technologies combined, present or future.

\section*{Acknowledgements}
\label{sec:acknowledgments}

We thank Kate Scholberg and Alec Habig for providing us useful information concerning SNEWS and NO$\nu$A and Christopher S. Kochanek for granting us information on optical sky surveys. We are very grateful to David Lhuillier for fruitful discussions.\\
With the support of the Technische Universit\"{a}t M\"{u}nchen - Institute for Advanced Study, funded by the German Excellence Initiative and the European Union Seventh Framework Programme under grant agreement n$^{\circ}$ 291763.

\begin{appendices}
\section{Toy Monte-Carlo}
\label{section:toyMC}

In order to generate an important amount of supernova neutrino events, we oriented our choice toward a fast toy Monte-Carlo simulation rather than a full GEANT4~\citep{Agostinelli:2002hh}. It implies a good understanding of IBD kinematics and detection to only keep its most important tendencies, without losing too much information. The code is written in MATLAB and uses GEANT4 outputs.

\subsection{Event generator}
Our toy Monte Carlo generates a set of neutrinos picked in a supernova spectrum flux. The (neutron, positron) pairs momenta and directions are created according to kinematics exposed in Section~\ref{subsec:IBDkinematics}. Their respective detection vertices are generated using data from Figures~\ref{fig:ranges} and~\ref{fig:capture_n}.
In the end, our toy Monte Carlo supplies a visible energy as well as prompt and delayed vertices.

\subsubsection*{Neutrino generation}
Neutrinos energies $E_\nu$ are picked at random in a supernova spectrum among the ones presented in Section~\ref{sec:SN_neutrino_signature}. Depending on the analysis, the number of generated events can either be user-defined or computed with respect to a defined supernova distance and detector size. 

\subsubsection*{$(e^+,n)$ pair generation}
For each event, a positron angle is picked using probability density function defined in Section~\ref{subsec:IBDkinematics}. From this angle and the initial $E_\nu$, the positron momentum $p_e = \sqrt{E_e^2 - m_e^2}$ is defined using Eq.~\ref{eqn:Ee2Enu}. Considering protons to be at rest in the laboratory frame, the neutron momentum is computed through the $p_n = p_\nuebar - p_e $ relation.
Until now, the generation of the $(e^+,n)$ pair was purely dictated by analytical IBD kinematics. However in order to obtain realistic prompt and delayed events, it is mandatory to simulate the behavior of the positron and neutron in a liquid scintillator. For simplicity reasons and since it is the current detector expected to detect the most neutrino events from a nearby supernova with the best sensitivity, we carried out simulations using the KamLAND liquid scintillator composed of a mixture of 80\% of C$_{12}$H$_{20}$ and 20\% of C$_{9}$H$_{12}$ with a density of 0.78 g/cm$^3$. After cross-checking the results using another liquid scintillator, we noticed no significant discrepancy between the two liquids and therefore used the KamLAND liquid scintillator for all simulations.

\subsubsection*{Prompt event}
Given their ionizing and penetrating nature, the path of positrons in the toy Monte Carlo is considered a straight line whose direction is given by the initial positron angle picked during the IBD generation. By simulating positrons with energies between 0 and 100~MeV in GEANT4, we obtained the mean positron range distribution with respect to the positron energy. In the toy Monte Carlo, the prompt event vertex is located on the positron direction and its distance from the interaction point is picked at random in the range distribution corresponding to its energy.
Note that after it deposits its energy along its pathway, the positron annihilates with an electron thus creating 2 gammas of total energy $2 m_{e}$, with $m_{e}$ the electron mass. Hence the energy deposition of the positron is not point-like but has a shape that depends on the initial positron energy. However, liquid scintillator detectors cannot resolve this shape and the reconstructed vertex usually corresponds to the barycenter of the charges. To efficiently take this effect into account in the toy Monte Carlo, we consider the path length of positrons from their creation to their annihilation and then apply a correction factor on the track accounting for point-like reconstruction of the positron track. Using GEANT4, the value of this correction factor was found to be equal to 0.72. The reconstructed prompt vertex is then smeared accordingly to the user-defined vertex resolution.

\subsubsection*{Delayed event}
As stated in Section~\ref{subsec:IBDkinematics}, the neutron loses its energy via elastically scattering off the nuclei composing the liquid, mainly light H nuclei. Given the small mass difference between a neutron and a H nucleus, neutrons lose half their kinetic energy, on average, at each scattering. Neutrons become rather quickly thermalized ($T_{n} \sim 0.025$~eV) and each scattering saves some directional information of the initial momentum, the first one being the most significant. After being thermalized, the neutron diffuses in the liquid before it gets captured on a H or Gd atom. This capture leads to the emission of gamma rays whose path distribution needs to be taken into account as well.
In our GEANT4 simulations, we first obtain the path length until thermalization for neutrons of different energies as displayed in Figure~\ref{fig:capture_n} (left). Then, by shooting thermal neutrons from this point, we obtain the diffusion path length before capture in an unloaded or a Gd-loaded liquid as seen in Figure~\ref{fig:capture_n} (right). Finally, in the case of a capture on H, we shot 2.2~MeV gamma rays to obtain their path length distribution. In the toy MC, the thermalization vertex is generated with respect to the mean path length depending on the neutron initial energy. The delayed vertex is the combination of the thermalization vertex and two isotropic vectors whose respective lengths are picked from the thermal neutron diffusion distribution and the gamma ray path length distribution. Like the prompt event, the reconstructed delayed vertex is smeared by the vertex resolution.

\begin{figure}[h!]
\begin{center}
\includegraphics[width = 0.6\textwidth]{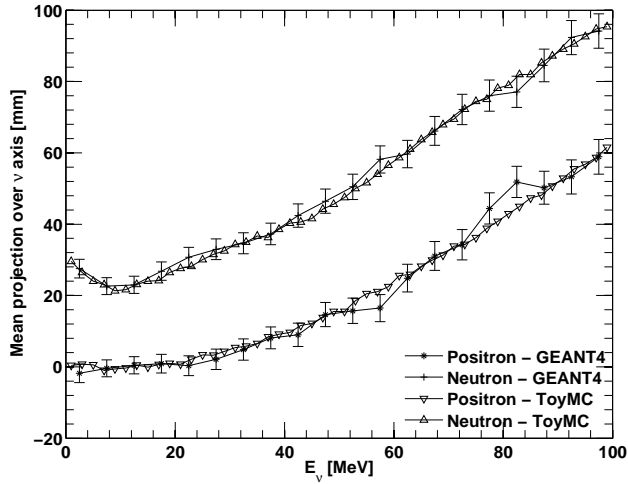}
\caption{Mean projections of prompt and delayed events along the $\nuebar$ direction obtained with a GEANT4 simulation and our tuned toy MC. The projections are expressed as a function of the neutrino energy.}
\label{fig:pr-del_projection_comparison}
\end{center}
\end{figure}

\subsection{Toy MC reliability}
\label{subsec:toyMCreliability}

To assess our toy MC IBD detection outline reliability, comparisons are made with GEANT4, for uniformly distributed neutrinos between 0 and 100~MeV, relevant energy for supernova neutrinos. 

The most relevant observables that can be used to compare the Delayed-Prompt vector in both GEANT4 and our toy MC are the prompt and delayed projections of the incoming neutrino axis. 
A 20~cm vertex resolution has been applied to all simulations. Figure~\ref{fig:pr-del_projection_comparison} shows the good agreement between our two simulation methods. 

\end{appendices}

\end{document}